\documentclass[12pt]{article}
\usepackage{axodraw}
\usepackage{epsf}

\def\theequation{\arabic{section}.\arabic{equation}}

\begin{document}

\begin{flushright}
   CERN-TH/2002-170\\[-0.15cm]
   MC-TH-2002-03\\[-0.15cm]
   hep-ph/0207277\\[-0.15cm]
   July 2002
\end{flushright}

\begin{center}
{\Large {\bf \hspace{-1cm}
Higgs-Mediated Electric Dipole Moments in the MSSM:\\[0.25cm] 
\hspace{-0.7cm} An Application to Baryogenesis and Higgs Searches}}\\[1.2cm]
{\large Apostolos Pilaftsis$^{\, a,b}$}\\[0.3cm] 
$^a${\em Theory Division, CERN, CH-1211 Geneva 23, Switzerland}\\
$^b${\em Department of Physics and Astronomy, University of Manchester,\\ 
Manchester M13 9PL, United Kingdom}
\end{center}
\bigskip \centerline{\bf ABSTRACT} 
{\small  We  perform a comprehensive study   of  the dominant two- and
higher-loop contributions to the $^{205}$Tl, neutron and muon electric
dipole moments  induced by Higgs bosons,   third-generation quarks and
squarks, charginos and gluinos in  the Minimal Supersymmetric Standard
Model~(MSSM).  We    find  that strong  correlations  exist  among the
contributing   CP-violating operators,  for   large  stop,  gluino and
chargino phases, and for  a wide  range  of values of $\tan\beta$  and
charged Higgs-boson  masses, giving rise to  large suppressions of the
$^{205}$Tl and   neutron electric dipole  moments below  their present
experimental   limits.  Based on  this    observation, we discuss  the
constraints that the non-observation of electric dipole moments imposes
on the  radiatively-generated CP-violating Higgs   sector and  on  the
mechanism  of electroweak   baryogenesis   in the  MSSM.   We  improve
previously suggested benchmark  scenarios of maximal  CP violation for
analyzing   direct  searches  of   CP-violating MSSM   Higgs bosons at
high-energy colliders, and  stress the important  complementary r\^ole
that a possible   high-sensitivity  measurement of the muon   electric
dipole moment  to the  level of  $10^{-24}$~$e$~cm   can play in  such
analyses. }


\newpage

\setcounter{equation}{0}
\section{Introduction}

The non-observation of electric dipole  moments (EDMs) of the thallium
atom  and neutron, as  well as the   absence of large flavour-changing
neutral-current (FCNC) decays put severe constraints on the parameters
of a theory.  Especially,  these  constraints become even more  severe
for supersymmetric theories, such as the MSSM, in which too large FCNC
and CP-violating effects  are  generically predicted at the   one-loop
level,   resulting  in gross violations  with    experimental data.  A
possible resolution  of such FCNC and CP  crises,  often considered in
the literature~\cite{GD}, makes  use  of the decoupling  properties of
the heavy  squarks and sleptons  of the  first two generations,  whose
masses should be larger than  $\sim$  10~TeV.  Thus, for  sufficiently
heavy squarks and sleptons, the one-loop  predictions for FCNC and EDM
observables can be suppressed up to levels compatible with experiment.
Also, such a solution poses no serious problem to the gauge hierarchy,
as long as the first  two generations of squarks  and sleptons are not
much heavier  than 10~TeV.  In this case,  because of their suppressed
Yukawa couplings, the radiative effect of the first two generations of
sfermions  on the Higgs-boson mass spectrum  is still negligible, with
respect  to that of TeV  scalar top and bottom quarks.\footnote[1]{One
should bear in mind that radiative  effects on the neutral Higgs-boson
masses are  proportional to the fourth  power of Yukawa  couplings.  A
simple  estimate  indicates  that  the   contribution of  the   second
generation of sfermions is smaller, by a factor of at least $10^{-7}$,
than those of the third generation.}

Recently,       it  has   been     shown~\cite{CKP,APedm}   that  even
third-generation squarks may lead by  themselves to observable effects
on the electron and neutron EDMs through Higgs-boson-mediated two-loop
graphs of the  Barr--Zee type~\cite{BZ}.  This observation  offers new
possibilities to  probe the   CP-violating soft-supersymmetry-breaking
parameters    related   to  the     third-generation squarks.    Most
interestingly, the same CP-violating parameters may induce radiatively
a  CP-noninvariant  Higgs-sector~\cite{APLB,PW,INch,CEPW}, leading  to
novel signatures at high-energy colliders~\cite{CEPW,CPX,CEPW2,CPHig}.
It is then obvious that EDM constraints do have important implications
for the phenomenological predictions within the above framework of the
MSSM with explicit CP violation.   Moreover, employing upper limits on
EDMs, one is, in principle, able to derive constraints on the phase of
the SU(2)$_L$ gaugino mass, $m_{\widetilde{W}}$, which plays a central
r\^ole      in      electroweak   baryogenesis~\cite{KRS}      in  the
MSSM~\cite{CQW,CMQSW}.

On the experimental side, the current  upper limit on the electron EDM
$d_e$, as derived   from the absence  of a  permanent atomic   EDM for
$^{205}$Tl, has improved by a  factor of  almost  2 over the last  few
years~\cite{EDMexp,EDMTl}. Specifically, the reported  $2\sigma$ upper
limit on a thallium EDM is~\cite{EDMTl}
\begin{equation}
  \label{dTlexp}
|d_{\rm Tl}|\ \stackrel{<}{{}_\sim}\ 1.3\times 10^{-24}~e\, {\rm cm}\, .
\end{equation}
Then, the electron EDM $d_e$ may be deduced indirectly by means of the
effective Lagrangian
\begin{eqnarray}
  \label{Lcp}
{\cal L}_{\rm EDM} &=& -\, \frac{1}{2}\, d_e\, \bar{e}\,
\sigma_{\mu\nu}\, i\gamma_5\,e\, F^{\mu\nu}\: +\:
C_S\, \bar{N}N\ \bar{e}\, i\gamma_5\,e\: +\: C_P\,  
\bar{N}\,i\gamma_5\,N\ \bar{e}\, e\nonumber\\
&&+\: C_T\,  \bar{N}\,
\sigma_{\mu\nu}\, i\gamma_5\,N\ \bar{e}\,\sigma^{\mu\nu}\, e\quad 
+\quad \dots\,,   
\end{eqnarray}
where  $C_S$,  $C_P$,  $C_T$  and  the  ellipses  denote  CP-violating
operators of  dimension 6  and higher. With  the aid of  the effective
Lagrangian (\ref{Lcp}),  the atomic EDM of $^{205}$Tl  may be computed
by~\cite{SBarr,FPT,KL}
\begin{eqnarray}
  \label{dTl}
d_{\rm Tl}\ [e\,{\rm cm}] &=& -\,585\times d_e\,[e\,{\rm cm}]\
+\ 8.5\times 10^{-19}\, [e\,{\rm cm}]\times C_S\, [{\rm TeV}^{-2}]\nonumber\\
&&-\, 8.\times 10^{-22}\, [e\,{\rm cm}]\times C_T\, [{\rm TeV}^{-2}]\ 
+\ \dots\, .
\end{eqnarray}
In (\ref{dTl}), the  dots denote CP-odd operators  of  dimension 7 and
higher.  In our analysis,  we will assume  that like $C_T$, the CP-odd
operators of   dimension  7  and  higher  give rise    generically  to
negligible  effects  on the   $^{205}$Tl~EDM.  Moreover, although  the
contributions of the neglected  CP-odd operators to other  heavy atoms
may be comparable to that of  $d_e$, the experimental upper limits are
still   much weaker than  $d_{\rm   Tl}$,  by  at least  one  order of
magnitude.  Consequently,  we  will only  analyze  predictions for the
thallium  EDM  $d_{\rm Tl}$  and   consider only  two  operators:  the
electron EDM  $d_e$  and the  CP-odd electron--nucleon  operator $C_S$.
{}From (\ref{dTlexp}) and  (\ref{dTl}),  it is  then  not difficult to
deduce the  following  $2\sigma$  upper limits   on  these two  CP-odd
operators:
\begin{equation}
  \label{deCSexp}
|d_e|\ \stackrel{<}{{}_\sim}\ 2.2\times 10^{-27}~e\,{\rm cm}\,,\qquad
|C_S|\ \stackrel{<}{{}_\sim}\  1.5\times 10^{-6}~[{\rm TeV}^{-2}]\, .
\end{equation}
In the  MSSM under  study, the contributions  from $d_e$ and  $C_S$ to
$d_{\rm Tl}$ can be of comparable size and therefore cannot be treated
independently.   In fact, depending  on their  relative sign,  one may
increase or  reduce the EDM  bounds on the CP-violating  parameters of
the  theory. Here,  the proposed  high-sensitivity measurement  of the
muon  EDM  $d_\mu$ to  the  level $10^{-24}$~$e$~cm~\cite{Yannis}  may
offer new constraints complementary to those obtained by $d_{\rm Tl}$,
since $C_S$ and all higher-dimensional CP-odd operators are absent.

Unlike the thallium  EDM, the upper limit on the  neutron EDM $d_n$ is
less severe,  i.e.\
\begin{equation}
  \label{dnexp}
|d_n|\ \stackrel{<}{{}_\sim}\  1.2\times 10^{-25}~e\,{\rm  cm}\,,
\end{equation}
at the  $2\sigma$ confidence level (CL)~\cite{dn90,dn99,LG}. Moreover,
although  promising   computations based  on   QCD sum rules~\cite{PR}
appeared recently,  the  theoretical prediction  for $d_n$  is  rather
model-dependent.  For     example,  the  predictions     between   the
valence-quark and quark-parton models may differ, even up to one order
of magnitude~\cite{AKL}.  Recently, the  experimental upper limit on a
permanent EDM of the $^{199}$Hg atom has been  improved by a factor of
4,  i.e.\ $|d_{\rm Hg}| <  2.33\times  10^{-28}$~$e\, {\rm cm}$ at the
$2\sigma$ CL~\cite{dHg}.    On  the  theoretical  side,  however,  the
derivation  of     bounds~\cite{FOPR}  from   $d_{\rm    Hg}$  on  the
chromoelectric dipole moment (CEDM) operators  of  $u$ and $d$  quarks
contains   many  uncertainties    related   to unknown   effects    of
higher-dimensional        chiral      operators,       nucleon-current
ambiguities~\cite{Ritz},  the   neglect   of  the  CP-odd  three-gluon
operator~\cite{SW,DDLPD}, the   modelling for extracting  the  nuclear
Schiff moment~\cite{KL}, the $s$-quark  content in heavy  nuclei, etc.
Thus, we shall not implement mercury EDM  constraints in our analysis.
Instead, we will consider  that no large cancellations~\cite{IN} below
the 10\% level occur among the different EDM terms in the neutron EDM.
In  a    sense,  such a   procedure   takes  account  of    a possible
complementarity relation~\cite{FOPR} between  the  measurements of the
neutron and Hg EDMs.

As  we  have  already  mentioned  above,  in  the  MSSM  one-loop  EDM
effects~\cite{EFN,PN,EDMrecent,AKL}  can be  greatly  suppressed below
their experimental limits, if the  first two generation of squarks and
sleptons   are    made   heavy   enough,    typically   heavier   than
10~TeV~\cite{PN,AKL}.  Within such  a framework of the MSSM~\cite{TI},
the   dominant  contributions  to   EDMs  arise   from  Higgs-mediated
Barr--Zee-type two-loop graphs that  involve quarks and squarks of the
third  generation,  charginos and  gluinos.\footnote[3]{Alternatively,
  one-loop EDM contributions can be suppressed if the CP phases of the
  trilinear soft-Yukawa couplings of the first two generations and the
  CP  phases   of  Wino  $\tilde{W}$,  Bino   $\tilde{B}$  and  gluino
  $\tilde{g}$  are all  zero,  with $B\mu$  and  $\mu$ being  positive
  according  to our  CP conventions.   In this  case, however,  if the
  first two  generations of sfermions are relatively  light, e.g.\ few
  hundreds  of GeV,  then additional  two-loop  EDM graphs~\cite{APRD}
  exist,  such  as  those  induced   by  a  gluino  CEDM,  which  give
  non-negligible  contributions to the  EDMs.  Furthermore,  there are
  two-loop EDM  effects induced by a CP-odd  $\gamma W^+W^-$ operator,
  which  do not  decouple in  the limit  of heavy  squarks and  do not
  depend  on   Higgs-boson  masses~\cite{THW}.   These   two-loop  EDM
  contributions  are  subdominant,   yielding  an  electron  EDM  term
  typically smaller  than $10^{-27}$~$e\, {\rm cm}$.}   In this paper,
we improve  previous computations  of these two-loop  contributions to
EDMs,  by  resumming  CP-even  and  CP-odd radiative  effects  on  the
Higgs-boson  self-energies and  vertices~\cite{CPH,Mhiggs}.  Analogous
improvements of  higher-order resummation effects  are also considered
in  the computation of  the CP-odd  electron--nucleon operator  $C_S$. 
Within  the  above  resummation  approach,  we  compute  the  original
Barr--Zee  EDM  graph  induced   by  $t$-quarks  beyond  the  two-loop
approximation  in  the MSSM  through  one-loop CP-violating  threshold
corrections to the top-quark  Yukawa coupling. Finally, we compute the
Higgs-boson  two-loop contribution  to EDM  induced by  charginos, and
discuss the consequences of the derived EDM constraints on electroweak
baryogenesis in the MSSM.

The present  paper is organized as  follows: in Section  2, we discuss
the CP-odd  electron--nucleon operator  $C_S$, which gives  an enhanced
contribution to the thallium EDM $d_{\rm Tl}$ in the large $\tan\beta$
regime~\cite{LP}.  In Section~3, after reviewing the existing dominant
Higgs-boson  two-loop  contributions  to  EDMs, we  compute  the  very
relevant Barr--Zee  contribution to EDM  from $t$ quarks for  the first
time in the MSSM.  In addition, we critically re-examine a very recent
calculation~\cite{CCK}  on  Higgs-boson two-loop  EDM  effects due  to
charginos.  In Sections~2 and~3, we also improve previous computations
of  the CP-odd electron--nucleon  operator $C_S$  and the  electron EDM
$d_e$, by taking properly into account higher-order CP-even and CP-odd
resummation  effects of  Higgs-boson  self-energy and  vertex graphs.  
Section~4 is devoted to numerical  estimates of EDMs and discusses the
impact of the derived  EDM constraints on electroweak baryogenesis and
on the analysis of direct searches for CP-violating Higgs bosons.  Our
conclusions are drawn in Section 5.

\setcounter{equation}{0}
\section{CP-odd electron--nucleon operator $C_S$}

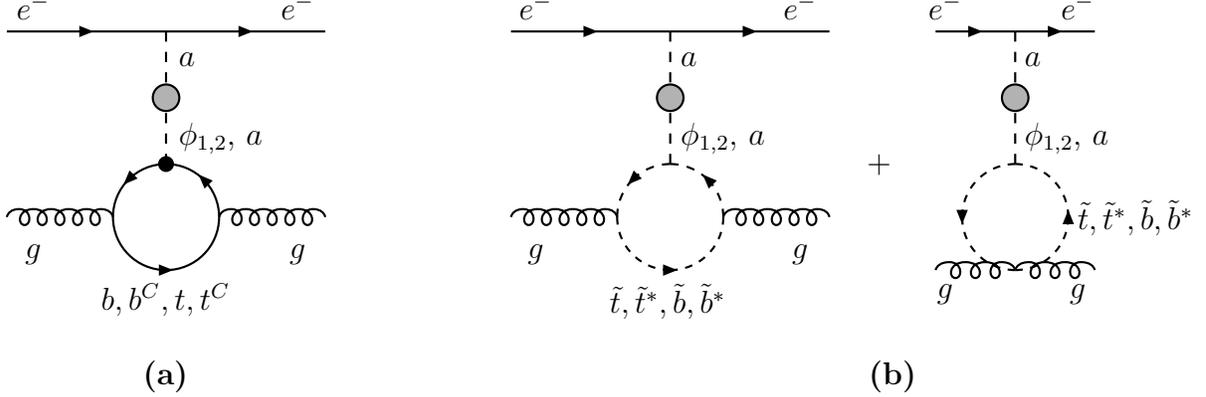
\begin{figure}

\begin{center}
\begin{picture}(450,150)(0,0)
\SetWidth{0.8}
 
\Gluon(10,50)(50,50){3}{5}\ArrowArc(70,50)(20,0,90)\ArrowArc(70,50)(20,90,180)
\ArrowArc(70,50)(20,180,360)\Gluon(90,50)(130,50){3}{5}
\DashLine(70,70)(70,120){4}
\ArrowLine(10,120)(70,120)\ArrowLine(70,120)(130,120)
\GCirc(70,95){5}{0.7}\Vertex(70,70){3}
\Text(75,110)[l]{$a$}\Text(75,80)[l]{$\phi_{1,2},\, a$}
\Text(20,125)[b]{$e^-$}\Text(120,125)[b]{$e^-$}
\Text(20,40)[t]{$g$}\Text(120,40)[t]{$g$}
\Text(70,25)[t]{$b,b^C,t,t^C$}

\Text(70,-10)[]{\bf (a)}

\Gluon(200,50)(240,50){3}{5}\DashArrowArc(260,50)(20,0,90){3}
\DashArrowArc(260,50)(20,90,180){3}
\DashArrowArc(260,50)(20,180,360){3}\Gluon(280,50)(320,50){3}{5}
\DashLine(260,70)(260,120){4}
\ArrowLine(200,120)(260,120)\ArrowLine(260,120)(320,120)
\GCirc(260,95){5}{0.7}
\Text(265,110)[l]{$a$}\Text(265,80)[l]{$\phi_{1,2},\, a$}
\Text(210,125)[b]{$e^-$}\Text(310,125)[b]{$e^-$}
\Text(210,40)[t]{$g$}\Text(310,40)[t]{$g$}
\Text(260,25)[t]{$\tilde{t},\tilde{t}^*,\tilde{b},\tilde{b}^*$}

\Text(340,70)[]{$+$}

\Gluon(360,30)(390,30){3}{3}\DashArrowArc(390,50)(20,-90,90){3}
\DashArrowArc(390,50)(20,90,270){3}\Gluon(390,30)(420,30){3}{3}
\DashLine(390,70)(390,120){4}
\ArrowLine(360,120)(390,120)\ArrowLine(390,120)(420,120)
\GCirc(390,95){5}{0.7}
\Text(395,110)[l]{$a$}\Text(395,80)[l]{$\phi_{1,2},\, a$}
\Text(365,125)[b]{$e^-$}\Text(415,125)[b]{$e^-$}
\Text(365,25)[t]{$g$}\Text(415,25)[t]{$g$}
\Text(415,50)[l]{$\tilde{t},\tilde{t}^*,\tilde{b},\tilde{b}^*$}

\Text(345,-10)[]{\bf (b)}

\end{picture}
\end{center}
\vspace{0.cm}
\caption{\em Feynman graphs contributing to a non-vanishing CP-odd
electron--nucleon operator $C_S$.  At the elementary particle level,
$C_S$ is predominantly induced by quantum effects involving
(a)~$t$-,$b$- quarks and (b)~$\tilde{t}$-, $\tilde{b}$- squarks. Blobs
and heavy dots denote resummation of self-energy and vertex graphs,
respectively.}\label{f1}
\end{figure}

Let  us first study  the contribution  of the  CP-odd electron--nucleon
operator  $C_S$~\cite{SBarr,FPT,KL}  to the  $^{205}$Tl  EDM.  At  the
elementary particle level, $C_S$ can be induced by two types of CP-odd
operators   in  supersymmetric   theories:   $\bar{e}  i\gamma_5   e\,
\bar{q}q$~\cite{LP} and $\bar{e} i\gamma_5 e\, \tilde{q}^* \tilde{q}$,
where   $q$  and   $\tilde{q}$   denote  quark   and  squark   fields,
respectively.   In  the  MSSM,  the  above  two  CP-odd  operators  of
dimensions  6  and 5  are  generated  by  interactions involving  Higgs
scalar--pseudoscalar mixing  and CP-violating vertex  effects, as those
shown in Fig.~\ref{f1}.

However, not all quarks and squarks can give rise to potentially large
contributions to the $^{205}$Tl EDM.  Our interest is to consider only
enhanced Yukawa and trilinear couplings of  the Higgs bosons to quarks
and squarks in  the decoupling limit   of the first  two generation of
squarks. This criterion singles  out the  CP-odd operators related  to
top and bottom quarks,  and their supersymmetric partners.  In fact,
as   is  shown in  Fig.~\ref{f1}, heavy   quarks  and  squarks  do not
contribute directly to the CP-odd operator $C_S$, but only through the
loop-induced Higgs--gluon--gluon   couplings $H_igg$, after  they have
been integrated out.   Thus, the effective Lagrangian responsible  for
generating $C_S$ is
\begin{equation}
  \label{Lcpeff}
{\cal L}^{(C_S)}_{\rm eff} \ =\  \sum_{i=1}^3\,
\frac{g_w\, H_i}{2M_W}\, \bigg(\, g_{H_igg}\,  \frac{\alpha_s}{8\pi}\, 
G^{a,\mu\nu}G^a_{\mu\nu}\ +\ m_e \tan\beta\ O_{3i}\, 
\bar{e}\,i\gamma_5\, e\, \bigg)\, ,
\end{equation}
where  $M_W =  g_w v/2$,  $O$ is  the $3\times  3$-mixing  matrix that
relates   the   weak  to   mass   eigenstates   of  the   CP-violating
Higgs bosons~\cite{PW,CEPW}, and
\begin{eqnarray}
  \label{Hgg}
g_{H_igg} \!&=&\! \sum\limits_{q=t,b}\, \bigg\{\,
\frac{2}{3}\ g^S_{H_iqq}\ +\ 
\frac{v^2}{6\, m^2_{\tilde{q}_1} m^2_{\tilde{q}_2}}\, 
\Big[\, (m^2_{\tilde{q}_2}
- m^2_{\tilde{q}_1})\, \xi^{(H_i)}_q\ +\ (m^2_{\tilde{q}_1} 
+ m^2_{\tilde{q}_2})\, \zeta^{(H_i)}_q\, \Big]\,\bigg\}\, .\qquad
\end{eqnarray}
In   (\ref{Hgg}),   the   dimensionless  coefficients   $g^S_{H_iqq}$,
$\xi^{(H_i)}_q$,  $\zeta^{(H_i)}_q$ and the  stop and  sbottom masses
are given in the appendix.

%******************************************************************
%%% Feynman graphs pertinent to the CP-odd four-fermion operator  
%******************************************************************

The largest contribution to  the coupling parameter $g_{H_igg}$  comes
from the  scalar  part of the  $H_i\bar{b}b$  coupling, $g^S_{H_ibb}$.
More  explicitly,  there  are two  CP-violating  effects that dominate
$g^S_{H_ibb}$:     (i)         the   $\tan^2\beta$-enhanced  threshold
effects~\cite{LP} described by the term
\begin{equation}
  \label{gSHbbtb}
g^S_{H_ibb} \ \sim\  {\rm Im}\, \bigg[\, \frac{(\Delta h_b/h_b)\, 
\tan^2\beta }{ 1\: +\: (\delta h_b/h_b)\: +\:  
(\Delta h_b/h_b)\, \tan\beta}\, \bigg]\, O_{3i}\, ,
\end{equation}
and  (ii)  the scalar--pseudoscalar  mixing  effects  contained in  the
mixing  matrix elements  $O_{1i}$.  The  definition of  the quantities
$\delta h_b/h_b$ and $\Delta h_b/h_b$ may be found in the appendix.

At this stage, it is important to observe  that if $(\Delta h_b/h_b)\,
\tan\beta  \stackrel{>}{{}_\sim} 1$,   the $\tan^2\beta$-dependence of
the CP-violating threshold   effects on $g^S_{H_i bb}$   and $g^P_{H_i
bb}$ considerably modifies.   In particular, in the large  $\tan\beta$
limit, $g^S_{H_i bb}$   and $g^P_{H_i bb}$  asymptotically  approach a
$\tan\beta$-independent constant, i.e.
\begin{eqnarray}
  \label{SHlim}
g^S_{H_i   bb} &\to &  {\rm Im}\,   \bigg[\,   \frac{1\:  +\: (\delta
h_b/h_b)}{(\Delta h_b / h_b)}\, \bigg]\, O_{3i}\, ,\\
  \label{PHlim}
g^P_{H_i   bb} &\to &  {\rm Im}\,   \bigg[\,   \frac{1\:  +\: (\delta
h_b/h_b)}{(\Delta h_b / h_b)}\, \bigg]\, O_{1i}\, .
\end{eqnarray}
Although   the above limits  may  only be  attainable  in a very large
$\tan\beta$ and quasi-nonperturbative regime  of the theory, the onset
of a $\tan\beta$-independent behaviour in $g^S_{H_i bb}$ and $g^P_{H_i
bb}$ may  already start from moderately  large  values of $\tan\beta$,
i.e.\  for $\tan\beta  \stackrel{>}{{}_\sim}  30$.   Consequently, the
limits  (\ref{SHlim})  and (\ref{PHlim}) should   be regarded as upper
bounds on the  CP-violating threshold-enhanced  parts of the  coupling
parameters $g^S_{H_i  bb}$  and  $g^P_{H_i  bb}$.  In  our   numerical
analysis   in Section~4,   we   properly    take  into  account    the
above-described CP-violating resummation effects on $g^S_{H_i bb}$.

The computation of the  CP-odd electron--nucleon operator $C_S$ can now
be performed  by utilizing standard  QCD techniques based on the trace
anomaly of the energy-momentum tensor~\cite{SVZ}.  In the chiral quark
mass limit, we then have the simple relation
\begin{equation}
  \label{qlimit}
\langle N |\, \frac{\alpha_s}{8\pi}\,
G^{a,\mu\nu}G^a_{\mu\nu}\, | N\rangle\ =\ -\, (100~{\rm MeV})\, \bar{N} N\,.
\end{equation}
With the help  of (\ref{qlimit}), we  can evaluate the  effective $H_i
\bar{N} N$ couplings, and hence the CP-odd operator $C_S$:
\begin{equation}
  \label{CSth} 
C_S\ =\ -\, (0.1~{\rm GeV})\, \tan\beta\ \frac{m_e \pi
\alpha_w}{M^2_W}\ \sum_{i=1}^3\, \frac{ g_{H_igg}\,
O_{3i}}{M^2_{H_i}}\ .
\end{equation}
Observe  that the operator  $C_S$  exhibits an enhanced  $\tan^3\beta$
dependence~\cite{LP};  it   therefore  becomes very   significant  for
intermediate and large values of $\tan\beta$.  Numerical estimates for
this contribution to a thallium EDM will be presented in Section~4.

\setcounter{equation}{0}
\section{Higgs-boson two-loop contributions to $d_e$}

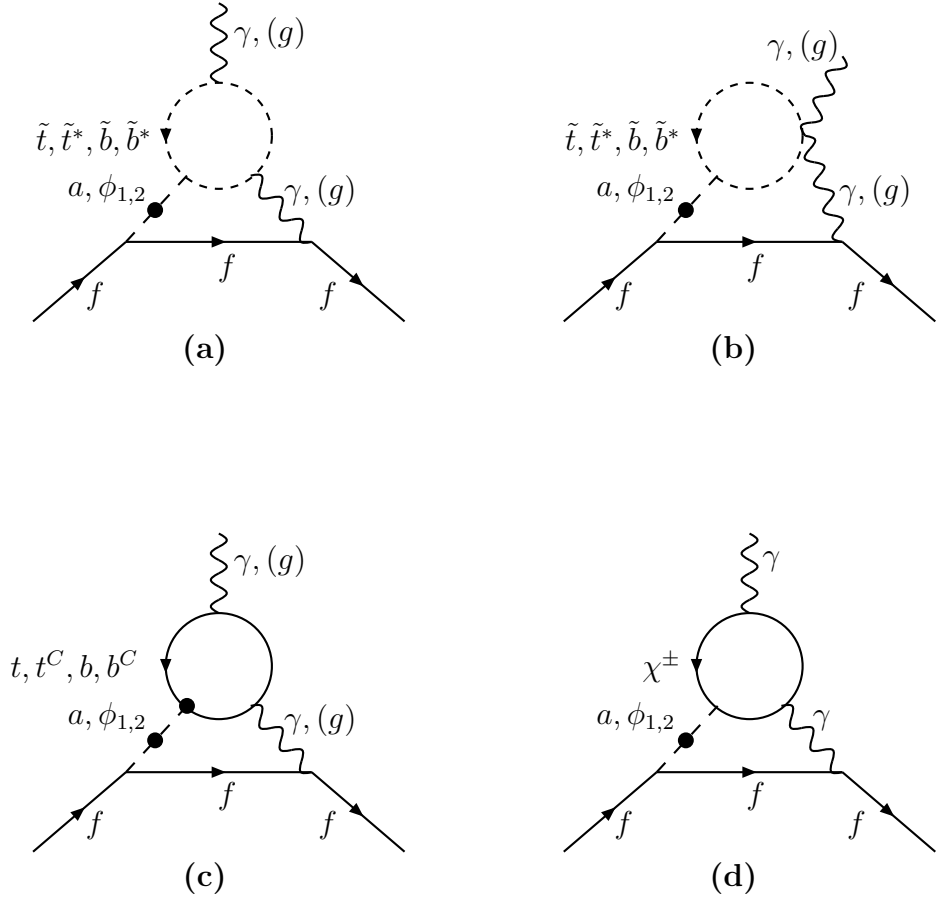
\begin{figure}
 
\begin{center}
\begin{picture}(400,400)(0,0)
\SetWidth{0.8}
 
\ArrowLine(10,230)(45,260)\Text(30,235)[lb]{$f$}
\DashLine(45,260)(68,285){5}\Vertex(56,272){3}
\Text(53,275)[rb]{$a, \phi_{1,2}$}
\Photon(92,285)(115,260){3}{3}\Text(105,280)[l]{$\gamma,(g)$}
\ArrowLine(45,260)(115,260)\Text(80,251)[l]{$f$}
\ArrowLine(115,260)(150,230)\Text(118,235)[lb]{$f$}
\Photon(80,320)(80,350){3}{3}\Text(86,340)[l]{$\gamma,(g)$}
\DashArrowArc(80,300)(20,0,360){3}
\Text(55,300)[r]{$\tilde{t},\tilde{t}^*,\tilde{b},\tilde{b}^*$} 

\Text(75,220)[]{\bf (a)}

\ArrowLine(210,230)(245,260)\Text(230,235)[lb]{$f$}
\DashLine(245,260)(268,285){5}\Vertex(256,272){3}
\Text(253,275)[rb]{$a, \phi_{1,2}$}
\Photon(301,300)(315,260){3}{4}\Text(315,280)[l]{$\gamma,(g)$}
\Photon(301,300)(315,330){3}{3}\Text(315,335)[r]{$\gamma,(g)$}
\ArrowLine(245,260)(315,260)\Text(280,251)[l]{$f$}
\ArrowLine(315,260)(350,230)\Text(318,235)[lb]{$f$}
\DashArrowArc(280,300)(20,0,360){3}
\Text(255,300)[r]{$\tilde{t},\tilde{t}^*,\tilde{b},\tilde{b}^*$} 

\Text(275,220)[]{\bf (b)}

\ArrowLine(10,30)(45,60)\Text(30,35)[lb]{$f$}
\DashLine(45,60)(68,85){5}\Vertex(68,85){3}\Vertex(56,72){3}
\Text(53,75)[rb]{$a, \phi_{1,2}$}
\Photon(92,85)(115,60){3}{3}\Text(105,80)[l]{$\gamma,(g)$}
\ArrowLine(45,60)(115,60)\Text(80,51)[l]{$f$}
\ArrowLine(115,60)(150,30)\Text(118,35)[lb]{$f$}
\Photon(80,120)(80,150){3}{3}\Text(86,140)[l]{$\gamma,(g)$}
\ArrowArc(80,100)(20,0,360)
\Text(50,100)[r]{$t,t^C,b,b^C$} 

\Text(75,20)[]{\bf (c)}

\ArrowLine(210,30)(245,60)\Text(230,35)[lb]{$f$}
\DashLine(245,60)(268,85){5}\Vertex(256,72){3}
\Text(253,75)[rb]{$a, \phi_{1,2}$}
\Photon(292,85)(315,60){3}{3}\Text(305,80)[l]{$\gamma$}
\ArrowLine(245,60)(315,60)\Text(280,51)[l]{$f$}
\ArrowLine(315,60)(350,30)\Text(318,35)[lb]{$f$}
\Photon(280,120)(280,150){3}{3}\Text(286,140)[l]{$\gamma$}
\ArrowArc(280,100)(20,0,360)
\Text(255,100)[r]{$\chi^\pm$} 

\Text(275,20)[]{\bf (d)}

\end{picture}
\end{center}
\caption{\em Dominant Higgs-boson two-loop contributions to EDM of a light
  fermion $f = e,\mu,d$ in the MSSM with explicit CP violation
  (mirror-symmetric graphs are not displayed). Heavy dots indicate
  resummation of self-energy and vertex graphs.  Two-loop graphs
  generating a CEDM for a $d$-quark are also shown.}\label{fedm}
\end{figure}

We  now       turn   our  attention     to     Higgs-boson    two-loop
effects~\cite{CKP,APedm} on the electron EDM  analogous to those first
discussed by  Barr and  Zee~\cite{BZ} in non-supersymmetric  theories.
As  is shown in  Fig.~\ref{fedm}, these two-loop EDM effects originate
predominantly from graphs   that  involve: stop and  sbottom   squarks
(Fig.~\ref{fedm}(a,b))~\cite{CKP},   top  and    bottom         quarks
(Fig.~\ref{fedm}(c)), and charginos (Fig.~\ref{fedm}(d))~\cite{CCK}.

%******************************************************************
%%% Higgs-boson two-loop contributions to EDM  
%******************************************************************

Strictly speaking,  the original Barr--Zee  graphs induced by  top and
bottom  quarks  in  Fig.~\ref{fedm}(c)  appear  beyond  the  two-loop
approximation  in the  MSSM.   However,  it is  a  formidable task  to
analytically  compute the  complete  set of  the  relevant three-  and
higher-loop  graphs.    Therefore,  we  consider  only   a  subset  of
higher-loop corrections,  in which the dominant  CP-violating terms in
the  Higgs-boson propagators  and the  Higgs-quark-quark  vertices are
resummed.  Such an approach should only be viewed as an effective one,
which  is  expected to  capture  the  main  bulk of  the  higher-order
effects.   In  the  same   vein,  we  improve  previous  two-loop  EDM
calculations  related  to   third-generation  squarks~\cite{CKP}  and
charginos~\cite{CCK}  by resumming  dominant  CP-violating self-energy
terms in the Higgs-boson propagators.

In  the  context  of  the  aforementioned  resummation  approach,  the
dominant  Higgs-boson  two-loop  contributions  to  electron  EDM  are
individually found to be
\begin{eqnarray}
  \label{2stsb}
\bigg(\frac{d_e}{e}\bigg)_{\rm (a,b)} \!\!\!\! & =& 
\frac{3\,\alpha_{\rm em}}{32\,\pi^3}\, m_e\, 
\sum_{i=1}^3\, \frac{g^P_{H_i ee}}{M^2_{H_i}}\,
\sum\limits_{q=t,b}\,  Q^2_q\, \bigg\{\, \xi^{(H_i)}_q\, 
\bigg[\, F\bigg(\frac{m^2_{\tilde{q}_1}}{M^2_{H_i}}\bigg)\ -\ 
F\bigg(\frac{m^2_{\tilde{q}_2}}{M^2_{H_i}}\bigg)\, \bigg]\nonumber\\
&&+\, \zeta^{(H_i)}_q\, 
\bigg[\, F\bigg(\frac{m^2_{\tilde{q}_1}}{M^2_{H_i}}\bigg)\ +\ 
F\bigg(\frac{m^2_{\tilde{q}_2}}{M^2_{H_i}}\bigg)\, \bigg]\, \bigg\}\,,\\
  \label{2quark}
\bigg(\frac{d_e}{e}\bigg)_{\rm (c)} & =& -\, 
\frac{3\,\alpha^2_{\rm em}}{8\,\pi^2\,\sin^2\theta_w}\
\frac{m_e}{M^2_W}\nonumber\\
&&\times \sum_{i=1}^3\ \sum\limits_{q=t,b}\, Q^2_q\, 
\bigg[\, g^P_{H_iee}\, g^S_{H_i qq}\, f\bigg(\frac{m^2_q}{M^2_{H_i}}\bigg)\ 
+\ g^S_{H_iee}\, g^P_{H_i qq}\, g\bigg(\frac{m^2_q}{M^2_{H_i}}\bigg)\,
\bigg]\,,\\
  \label{2chargino}
\bigg(\frac{d_e}{e}\bigg)_{\rm (d)} & =& -\, 
\frac{\alpha^2_{\rm em}}{8\,\sqrt{2}\,\pi^2\,\sin^2\theta_w}\
\frac{m_e}{M_W}\nonumber\\
&&
\times \sum_{i=1}^3\ \sum\limits_{j=1,2}\, \frac{1}{m_{\chi^+_j}}\, 
\Bigg[\, g^P_{H_iee}\, a_{H_i \chi^-_j\chi^+_j}\, 
f\Bigg(\frac{m^2_{\chi^+_j}}{M^2_{H_i}}\Bigg)\ 
+\ g^S_{H_iee}\, b_{H_i \chi^-_j \chi^+_j}\, 
g\Bigg(\frac{m^2_{\chi^+_j}}{M^2_{H_i}}\Bigg)\,\Bigg] ,\qquad
\end{eqnarray}
where $g^P_{H_i ee} = -\, \tan\beta\,  O_{3i}$, $g^S_{H_i ee} = O_{1i}
/ \cos\beta$, and
\begin{eqnarray}
  \label{FZ}
F(z) &=& \int_0^{1} dx\ \frac{x(1-x)}{z\: -\: x(1-x)}\ 
\ln \bigg[\,\frac{x(1-x)}{z}\,\bigg]\, ,\\
  \label{fz}
f(z) &=& \frac{z}{2}\, \int_0^{1} dx\ \frac{1\: -\: 2x(1-x)}{x(1-x)\: -\: z}\ 
\ln \bigg[\,\frac{x(1-x)}{z}\,\bigg]\, ,\\
  \label{gz}
g(z) &=& \frac{z}{2}\, \int_0^{1} dx\ \frac{1}{x(1-x)\: -\: z}\ 
\ln \bigg[\,\frac{x(1-x)}{z}\,\bigg]\, 
\end{eqnarray}
are  two-loop  functions.   The  coupling  coefficients $g^S_{H_iqq}$,
$g^P_{H_iqq}$,  $a_{H_i    \chi^-_j\chi^+_j}$ and   $b_{H_i   \chi^-_j
\chi^+_j}$ in (\ref{2stsb})--(\ref{2chargino}),  as well as the squark
and     chargino    masses,     are    given    in    the    appendix.
Equation~(\ref{2stsb}) takes on  the  simpler analytic  form presented
in~\cite{CKP}, if only the CP-odd component $a$ of the Higgs bosons is
considered in an unresummed two-loop calculation  of the EDM.  In this
case, the   coefficients $\zeta^{(H_i)}$  vanish   and $\xi^{(H_i)}_q$
simplifies to
\begin{equation}
\xi_q\ =\ R_q\, \frac{\sin 2\theta_q m_q\, {\rm Im}\, ( \mu
             e^{i\delta_q})}{\sin\beta\, \cos\beta\, v^2}\ =\
             \frac{R_q}{\sin\beta\cos\beta}\ \frac{2m^2_q\, {\rm
             Im}\,(\mu A_q)}{v^2\, (m^2_{\tilde{q}_2} -
             m^2_{\tilde{q}_1})}\ ,
\end{equation}
where $\delta_q = {\rm arg}  (A_q - R_q \mu^*)$,  with $R_t\, (R_b)  =
\cot\beta\ (\tan\beta)$.

In addition  to the dominant Higgs-boson  two-loop graphs we have been
studying here, there are also subdominant two-loop EDM diagrams, where
the virtual  photon is  replaced by  a  $Z$ boson  in Fig.~\ref{fedm}.
Another class of Higgs-boson two-loop  graphs involve the coupling  of
the  charged Higgs  bosons  $H^\pm$  to the  photon   and the  $W^\mp$
bosons~\cite{APedm}.  In  this  case,  for   example, the    graph  in
Fig.~\ref{fedm}  will   proceed via  charginos  and neutralino  in the
fermionic loop. As has  been explicitly shown in~\cite{APedm} for most
of the cases,  this additional set of  graphs give almost one order of
magnitude smaller contributions   to  EDM.  Most importantly,    their
dependence  on the  CP-violating  parameters of the  theory  is rather
closely    related    to   the  two-loop     EDM  graphs   depicted in
Fig.~\ref{fedm}.  Thus,  suppressing the dominant Higgs-boson two-loop
contributions   to  EDM  will automatically  lead  to  a corresponding
suppression of this additional  set of two-loop graphs.  Therefore, in
our analysis we neglect the aforementioned set of subdominant two-loop
graphs.

So  far, we  have  only been  studying  the electron  EDM $d_e$.   The
two-loop prediction  for the muon  EDM $d_\mu$ can easily  be obtained
from $d_e$ by considering the obvious mass rescaling factor $m_\mu/m_e
\approx 205$, i.e.
\begin{equation}
  \label{dmu}
d_\mu\ \approx\ 205\, d_e (t,b,\tilde{t},\tilde{b},\chi^\pm)\, ,
\end{equation}
where the different   two-loop EDM contributions  are indicated within
the parentheses.

On  the other hand, the  dominant  contributions to  neutron EDM $d_n$
come    from the  CEDM  of  the   $d$   quark and  CP-odd  three-gluon
operator~\cite{DDLPD}, which was first discussed by Weinberg~\cite{SW}
in non-supersymmetric multi-Higgs doublet   models.  In the  MSSM, the
CP-odd   three-gluon  operator  decouples   as $1/m^3_{\tilde{g}}$ and
becomes  relevant at gluino masses below  the TeV  scale.  The CEDM of
the $d$  quark may be obtained  from (\ref{2stsb}) and (\ref{2quark}),
if one replaces  the colour factor 3  by  1/2, and $\alpha_{\rm  em}\,
Q^2_q$ by  $\alpha_s$.    The  computation of the   neutron  EDM $d_n$
involves a   number  of  hadronic uncertainties,  when   the  EDMs are
translated  from  the  quark  to   the hadron  level~\cite{AKL}.   For
example, considering the valence-quark model and renormalization-group
running effects  from    the  electroweak  scale  $M_Z$ down   to  the
low-energy hadronic scale $\Lambda_h$~\cite{APedm}, one may be able to
establish an approximate  relation between neutron  and electron EDMs.
Thus,  taking the input  values for the involved kinematic parameters:
$m_d   (\Lambda_h)  =  10$~MeV, $\alpha_s   (M_Z)   =  0.12$ and  $g_s
(\Lambda_h)/(4\pi) = 1/\sqrt{6}$, we find
\begin{equation}
  \label{dn}
d_n\ \approx\ -10\, d_e (t,\tilde{t}) \: +\: 1.2\, 
d_e (t,\tilde{t},\chi^\pm)\: +\: d^{3G}_n\, .
\end{equation}
On  the RHS of (\ref{dn}),  the  first and  second  terms arise from a
$d$-quark   CEDM and   EDM, respectively,    and  $d^{3G}_n$  is   the
contribution  to  $d_n$ due to  the CP-odd  three-gluon  operator.  In
obtaining (\ref{dn}), we  have  made two additional  approximations as
well.  First, we neglected the contribution of the $u$-quark EDM $d_u$
to $d_n$, as it is  much smaller than the $d$-quark  EDM $d_d$ for the
relevant  region   $\tan\beta \stackrel{>}{{}_\sim}  3$.    Second, we
ignored the $b$- and  $\tilde{b}$-quantum corrections to $d_d$  and so
to $d_n$.  Formulae~(\ref{dmu}) and~(\ref{dn})  will be used to obtain
numerical  predictions  for the  muon  and neutron   EDMs in  the next
section.

\setcounter{equation}{0}
\section{Numerical estimates and discussion}

In Sections 2    and 3, we  computed the   dominant two-loop  and  the
resummed higher-loop    contributions to   EDMs  that originate   from
third-generation quarks  and  squarks,  and charginos.  Based   on the
derived analytic expressions,  we   can now analyze  numerically   the
impact of the experimental  constraints due to the  non-observation of
thallium  and  neutron  EDMs on the    CP-violating parameters of  the
theory, and hence on  electroweak baryogenesis and direct searches for
CP-violating   Higgs bosons in  the   MSSM.  Moreover, we will present
predictions for the muon EDM $d_\mu$ and discuss the implications of a
possible high-sensitivity    measurement of   $d_\mu$  to  the   level
$10^{-24}$~$e$~cm for our analyses.

Based on   the observation that CP-violating   quantum effects  on the
neutral Higgs sector get enhanced   when the product ${\rm Im}\,  (\mu
A_t)/    M^2_{\rm  SUSY}$  is    large~\cite{APLB,PW},  the    authors
in~\cite{CEPW,CPX} introduced  a  benchmark scenario,   called CPX, in
which  the   effects of  CP violation    are maximized.  In   CPX, the
following values for the $\mu$- and soft-SUSY-breaking parameters were
adopted:
\begin{eqnarray}
  \label{CPX}
\widetilde{M}_Q \!&=&\! \widetilde{M}_t\ =\ 
\widetilde{M}_b\ =\ M_{\rm SUSY}\,,\qquad
\mu \ =\ 4 M_{\rm SUSY}\,,\nonumber\\
|A_t| \!&=&\!  |A_b|\ =\ 2M_{\rm SUSY}\,,\qquad 
{\rm arg}(A_{t,b})\ =\ 90^\circ\,, \nonumber\\   
|m_{\tilde{g}}| \!&=&\! 1~{\rm TeV}\,,\qquad 
{\rm arg}(m_{\tilde{g}})\ =\ 90^\circ\,,\nonumber\\
m_{\widetilde{W}} \!&=&\! m_{\widetilde{B}}\ =\ 0.3~{\rm TeV}\, .
\end{eqnarray}
Without loss of generality, the $\mu$-parameter is  chosen to be real.
The predictions   of CPX showed~\cite{CPX}  that even  a light neutral
Higgs boson with a mass as low as  60~GeV could have escaped detection
at  LEP2.\footnote[3]{Similar remarks were  made earlier in~\cite{PW},
but the LEP2 data were less  restrictive then.}  A recent experimental
analysis of LEP2 data confirms this observation~\cite{OPAL}.  Here, we
wish  to investigate  the compatibility of   the CPX scenario with the
experimental limits on EDMs.  For this purpose, we allow variations in
the gluino phase, which enters the Higgs sector at two loops, but keep
the $A_t$  phase in (\ref{CPX}) fixed.   In addition, we  will present
numerical  results for EDMs, where  the $\mu$-parameter is varied from
100~GeV to $4 M_{\rm SUSY}$.  Finally, we leave unspecified the phases
of  the   gaugino       mass  parameters   $m_{\widetilde{W}}$     and
$m_{\widetilde{B}}$.     As we   will     see  below, the   phase   of
$m_{\widetilde{W}}$  is   greatly  affected by   constraints from  the
electron EDM.

We start our numerical analysis by presenting predicted values for the
$^{205}$Tl~EDM $d_{\rm  Tl}$ that arise   entirely  due to the  CP-odd
electron--nucleon operator   $C_S$  and are   denoted  as  $d_{\rm Tl}
(C_S)$.  In Fig.~\ref{fig1},  we    display numerical estimates    for
$d_{\rm  Tl} (C_S)$ as   functions of $\tan\beta$  for four  different
versions of the CPX scenario with $M_{\rm SUSY} = 1$~TeV: (a) $M_{H^+}
=  150$~GeV, ${\rm arg}\,  (m_{\tilde{g}}) = 0^\circ$;  (b) $M_{H^+} =
300$~GeV, ${\rm  arg}\,  (m_{\tilde{g}})  = 0^\circ$;  (c)  $M_{H^+} =
150$~GeV,  ${\rm arg}\, (m_{\tilde{g}})   = 90^\circ$; (d) $M_{H^+}  =
300$~GeV, ${\rm arg}\,   (m_{\tilde{g}}) = 90^\circ$.   The individual
$b,\tilde{t},t,\tilde{b}$  contributions  to  $d_{\rm Tl}(C_S)$, along
with their relative signs, are indicated by  different types of lines.
We  observe that the largest  contribution to  $d_{\rm Tl}$ comes from
the  $b$-quarks for large values  of $\tan\beta$, i.e.\ for $\tan\beta
\stackrel{>}{{}_\sim} 15$, for which  the CP-violating  vertex effects
become  important   (see  also the  discussion     in Section 2).   In
particular,  these  CP-violating threshold   effects,  which crucially
depend  on the  term  ${\rm Im}\,  (\Delta h_b/h_b)\, \tan^2\beta$  in
(\ref{gSHbbtb}),  become even more  important for large gluino phases.
Thus, the predictions for  $d_{\rm Tl} (C_S)$  in panels~\ref{fig1}(a)
and (b), with ${\rm  arg}\, (m_{\tilde{g}}) =  0^\circ$, are one order
of magnitude larger  than the ones in (c)  and (d),  with ${\rm arg}\,
(m_{\tilde{g}}) = 90^\circ$.

For intermediate  and  smaller    values of  $\tan\beta$,  i.e.\   for
$\tan\beta \stackrel{<}{{}_\sim} 15$, CP-violating self-energy effects
are significant, especially for  relatively light charged Higgs bosons
with masses in  the range 150--200~GeV.   In fact, these effects  have
generically opposite sign to  the CP-violating vertex effects,  giving
rise to natural cancellations among the contributing EDM terms, and so
lead to smaller values of $d_{\rm Tl}  (C_S)$.  Although our numerical
results are in qualitative  agreement with those in Ref.~\cite{LP}, we
actually   find noticeable   quantitative  differences,  when resummed
CP-violating self-energy and vertex effects are considered at the same
time.

Next,   we  shall  investigate numerically  higher-order  CP-violating
vertex  and self-energy effects induced  by $t$- and $b$-quarks on the
electron EDM   $d_e$.  Fig.~\ref{fig2}  shows numerical  estimates for
those  resummed effects on   $d_e$  as functions  of  $\tan\beta$,  in
variants   of the  CPX  scenario,  with  (a)~$M_{H^+}  = 150$~GeV  and
(b)~$M_{H^+} = 300$~GeV.  In particular, we considered three different
choices of the gluino phase:  ${\rm arg} (m_{\tilde{g}} ) = 90^\circ,\
0,\ -90^\circ$, denoted  as  $t_{1,2,3}$, respectively.  We  find that
CP-violating threshold corrections to the $H_itt$ coupling as small as
5\% are sufficient to  lead to  observable  EDM values for  $d_e$.  In
this respect, we see that the $t$-quark effects strongly depend on the
gluino phase through the combination  ${\rm Im}\, (\mu m_{\tilde{g}})$
that occurs in ${\rm   Im}\,  (\Delta h_t/h_t)$ [cf.\   (\ref{gSHtt}),
(\ref{gPHtt})].  Thus, the $t$-quark contribution to $d_e$ is positive
(negative)  for negative  (positive) gluino phases,   while  it is one
order of magnitude smaller  and negative for vanishing gluino  phases,
i.e.\ for ${\rm arg}\, (m_{\tilde{g}}) = 0^\circ$.  For comparison, we
also  included in     Fig.~\ref{fig2}  the   dependence of    positive
stop/sbottom contributions to       $d_e$~\cite{CKP} (long-dash-dotted
lines)    on  $\tan\beta$.   The   sum of    the $t$-,   $b$-quark and
$\tilde{t}$-, $\tilde{b}$-squark contributions  to  $d_e$ is  given by
the  solid lines $1,2,3$  for the  same values  of  gluino phases.  As
before, we indicate negative contributions to $d_e$ with a minus sign.
{}From {}Figs.~\ref{fig2}(a) and (b), it is interesting to notice that
if ${\arg}\,  (m_{\tilde{g}}) =  90^\circ$   in CPX,  a   cancellation
between the  $t$-quark and $\tilde{t}$-squark EDM contributions occurs
for almost the entire  range of the perturbatively allowed $\tan\beta$
values and  for  all   phenomenologically viable charged   Higgs-boson
masses.  As  a consequence of  such a  cancellation, the  electron EDM
$d_e$ is always smaller than the  current $2\sigma$ experimental limit
on  $d_e$,  i.e.\ $d_e  <  2.2\times 10^{-27}$~$e$~cm, even  for large
$\tan\beta$ values up to 30. As we will see below, this prediction may
considerably change  if  contributions from the CP-violating  operator
$C_S$ or chargino two-loop effects are considered.

In order to further gauge the importance of the $t$-quark two-loop EDM
effects, we present  in Fig.~\ref{fig2new} numerical values  for $d_e$
versus the $\mu$-parameter  for $\tan\beta = 20$,  and for two charged
Higgs-boson   masses:  (a) $M_{H^+}   =  150$~GeV  and (b)  $M_{H^+} =
300$~GeV.   The soft-SUSY-breaking parameters   are chosen as given in
(\ref{CPX})   for $M_{\rm    SUSY}    =  1$~TeV,   except   for    the
$\mu$-parameter, which has  been varied  from  $0.1$--4~TeV.  For  the
sake  of comparison,  we  also included  the Higgs-boson  two-loop EDM
effects induced by $\tilde{t}$-  and $\tilde{b}$-squarks.  The meaning
of   the various types  of   lines is exactly   the  same as those  in
Fig.~\ref{fig2}.  Remarkably enough, we find that even $\mu$ values as
low  as 500~GeV may be  sufficient to lead to  an electron  EDM at the
observable level through    the original two-loop  Barr--Zee  graph in
Fig.~\ref{fedm}(c).   In this  context,   we  also observe  that   the
resummed Higgs-boson two-loop  contributions to  $d_e$ from $t$-quarks
are  comparable   and    even   larger   than  those   coming     from
$\tilde{t}$-squarks for maximal gluino phases.  In fact, if $A_{t,b} =
0$,   the $\tilde{t}$-squark  and  dominant  CP-violating Higgs-mixing
effects  may be completely  switched  off, without much affecting  the
corresponding $t$-quark two-loop contributions to $d_e$.  Note that in
this case the $t$-quark effects on  $d_e$ and the $b$-quark effects on
the $C_S$  operator, which both formally arise  at the two-loop level,
are proportional to ${\rm Im}\, (\mu m_{\tilde{g}})$.  Therefore, they
turn out to be strongly correlated  and their combined contribution to
$d_{\rm  Tl}$ should carefully  be    taken  into account (see    also
discussions of Figs.~\ref{fig4} and \ref{fig5} below).

As was already  pointed out in~\cite{CKP,APedm}, charginos might  also
contribute to  electron EDM $d_e$  at the two-loop level.  Recently, a
computation  of  those  effects appeared  in~\cite{CCK}.   The authors
derived strict constraints  on    the CP-violating parameters    of  a
scenario in which electroweak baryogenesis is mediated by CP-violating
currents involving chargino interactions.    Here, we re-examine  this
issue   within  a   scenario that  favours   the   above  mechanism of
electroweak  baryogenesis and is not in  conflict  with LEP2 limits on
the   Higgs-boson   masses   and   couplings.    Specifically,   being
conservative, we require that  these be $M_{H_i} \stackrel{>}{{}_\sim}
111$~GeV,     for   $g^2_{H_iZZ} \stackrel{>}{{}_\sim}    0.3$,  where
$g_{H_iZZ}$  is the $H_iZZ$ coupling given  in units of the SM $h_{\rm
SM}   ZZ$ coupling.  In  addition, we  demand  that $M_{H_i} + M_{H_j}
\stackrel{>}{{}_\sim} 170$~GeV.  On   the  other hand, in  order   for
electroweak baryogenesis  to   proceed   via a   sufficiently   strong
first-order  phase  transition, the right-handed  stop  mass parameter
$\widetilde{M}_t$ must be  rather  small, and the  $\mu$  and the soft
gaugino parameter $m_{\widetilde{W}}$ must not be too large, typically
smaller  than 0.5~TeV~\cite{CQW,CMQSW}.     Especially,   there is   a
resonant  enhancement  up  to even   10  times  the   observed  baryon
asymmetry,   if  the    condition    $\mu  =    m_{\widetilde{W}}$  is
met~\cite{CMQSW}.  Further   requirements for  a scenario   leading to
successful electroweak   baryogenesis  are:  (i)~a  moderate trilinear
$A_t$-parameter     in       the      range, $0.2\stackrel{<}{{}_\sim}
A_t/\widetilde{M}_Q \stackrel{<}{{}_\sim} 0.65$; (ii) a not very large
$\tan\beta$ value,    $\tan\beta  \stackrel{<}{{}_\sim} 20$;   (iii) a
soft-SUSY-breaking   parameter $\widetilde{M}_Q$  of a  few   TeV, for
phenomenological reasons~\cite{CMQSW}.  More explicitly, the following
values for the mass parameters are employed:
\begin{eqnarray}
  \label{EWbau}
\widetilde{M}_Q \!&=&\! 3~{\rm TeV}\,, \qquad \widetilde{M}_t\ =\ 0\,,
\qquad  \widetilde{M}_b\ =\ 3~{\rm TeV}\,,\nonumber\\
|A_t| \!&=&\!  |A_b|\ =\ 1.8~{\rm TeV}\,,\qquad 
{\rm arg}(A_{t,b})\ =\ 0^\circ\,,\qquad \tan\beta\
\stackrel{<}{{}_\sim}\ 20\, ,  \nonumber\\   
|m_{\tilde{g}}| \!&=&\! 3~{\rm TeV}\,,\qquad 
{\rm arg}(m_{\tilde{g}})\ =\ 0^\circ\, ,\nonumber\\
\mu \!&=&\!  |m_{\widetilde{W}}|\ \stackrel{<}{{}_\sim}\ 
0.5~{\rm TeV}\,,\qquad 
{\rm arg} (m_{\widetilde{W}}) = 90^\circ\,.
\end{eqnarray}

To be able  to compare our predictions with  those presented in Fig.~2
of Ref.~\cite{CCK}, we choose in Fig.~\ref{fig3}(a) the values: $\mu =
m_{\widetilde{W}}   =  0.2$~TeV  and   $M_{H^+}  =   170$~GeV.   Since
CP-violating Higgs-mixing effects in the mass spectrum are generically
small for  the chosen values  of the parameters  in~(\ref{EWbau}), our
mass  input $M_{H^+} =  170$~GeV corresponds  to $M_{\mbox{\scriptsize
    `$A$'}} \approx 150$~GeV  for the mass of the  almost CP-odd Higgs
scalar $A$.   Even though  on a very  qualitative basis  our numerical
results on  the linear  $\tan\beta$-increase behaviour of  $d_e$ agree
with those  reported in~\cite{CCK}, the  actual functional dependences
of  the  individual `$h$',  `$H$',  `$A$'  contributions  to $d_e$  on
$\tan\beta$  differ  significantly.   Unlike~\cite{CCK},  we  find  in
Fig.~\ref{fig3}(a) that  for $\tan\beta \stackrel{>}{{}_\sim}  5$, the
$\tan\beta$-enhanced effect on $d_e$ originates from the heavier Higgs
bosons `$H$' and `$A$', while the EDM contribution due to the lightest
Higgs boson `$h$' is almost negligible.\footnote[4]{The fact that only
  `$H$' and  `$A$' contributions to $d_e$ exhibit  a linearly enhanced
  dependence on  $\tan\beta$ may also  be verified independently  by a
  flavour-graph  analysis.} Since  the size  of  $d_e$ is  set by  the
heavier  Higgs-boson masses,  i.e.\  by $M_{H^+}$,  and  by $\mu$  and
$m_{\widetilde{W}}$,  our  predictions  are  rather robust  under  the
different  choices  of the  remaining  soft-SUSY-breaking parameters.  
Moreover, although our numerical  values for the total contribution to
$d_e$  agree  very well  with~\cite{CCK}  for  $\tan\beta  = 2$  ($d_e
\approx 0.63\times  10^{-26}$~$e$~cm), they are  smaller by $\sim$20\%
for  $\tan\beta   =  6$,  i.e.\   we  find  $d_e   \approx  1.62\times
10^{-26}$~$e$~cm, which  should be compared with  $d_e \approx 2\times
10^{-26}$~$e$~cm.      Finally,    the     electroweak    baryogenesis
scenario~(\ref{EWbau})  in  the  low  $\tan\beta$  region,  $\tan\beta
\stackrel{<}{{}_\sim}   6$,   which   is   studied  by   the   authors
in~\cite{CCK}, appears to be highly disfavoured by LEP2 data.  In this
respect, a phenomenologically viable  model, with $M_{H^+} = 170$~GeV,
would   require  larger  values   of  $\tan\beta$,   i.e.\  $\tan\beta
\stackrel{>}{{}_\sim} 9$.  In this case,  one has to consider a factor
of  10 suppression  in  the  chargino phase,  such  that the  chargino
two-loop EDM effects are reduced  to a level close to the experimental
upper limit  on $d_e$.  Consequently, if no  cancellations are assumed
with  possible  one-loop  EDM  terms,  then a  model  with  suppressed
chargino phase of $\sim 5^\circ$  and a relatively light charged Higgs
boson, $M_{H^+}  = 150$--200~GeV, might  still be possible  to account
for  the  observed baryon  asymmetry  in  the  Universe, provided  the
aforementioned  resonant  factor  10  is  used.   However,  the  above
situation may be considerably  relaxed for larger values of $M_{H^+}$,
since   the  chargino   two-loop   EDM  effect   on  $d_e$   decreases
approximately by $1/M_{H^+}$  as $M_{H^+}$ increases.  This dependence
of $d_e$ on $M_{H^+}$ can explicitly be seen in the lower panel (b) of
Fig.~\ref{fig3}, for increasing charged Higgs-boson masses: $M_{H^+} =
150$~GeV~(solid),          200~GeV~(dashed),         300~GeV~(dotted),
500~GeV~(dash-dotted) and 1~TeV~(long-dash-dotted), in a scenario with
$m_{\widetilde{W}} = \mu = 0.4$~TeV and ${\rm arg} (m_{\widetilde{W}})
= 90^\circ$.

In the following, we  will present predictions  for more realistic EDM
observables,  with relatively  reduced hadronic  uncertainties, namely
the thallium EDM $d_{\rm  Tl}$, the neutron EDM $d_n$,  as well as the
muon EDM $d_\mu$  which  was  suggested  to  be measured with  a  high
sensitivity to  the   level  of  $10^{-24}$~$e$~cm~\cite{Yannis}.   In
Fig.~\ref{fig4}, we display numerical  values for $d_{\rm  Tl}$, $d_n$
and  $d_\mu$ as functions of $\tan\beta$   in two versions  of the CPX
scenario,  with $M_{H^+} =  150$~GeV: (a) ${\rm arg} (m_{\tilde{g}}) =
{\rm   arg}   (m_{\widetilde{W}} )  = 90^\circ$,  and   (b) ${\rm arg}
(m_{\tilde{g}}) =   35^\circ$,   ${\rm arg}  (m_{\widetilde{W}}   )  =
90^\circ$.   Fig.~\ref{fig4}  also  shows the different contributions,
along    with their  relative signs,    to $d_{\rm  Tl}$ from top/stop
(long-dash-dotted) and  chargino (dotted) Higgs-boson two-loop graphs,
as well as from the CP-odd  electron--nucleon operator $C_S$ (dashed).
Note that the type of lines used to represent the numerical results of
the individual   EDM contributions is given   in the  parentheses.  In
panel (a) of Fig.~\ref{fig4}, we   see that the contribution of  $C_S$
prevails in $d_{\rm Tl}$, for large values of $\tan\beta$, and exceeds
the experimental limit for $\tan\beta  \stackrel{>}{{}_\sim} 12$.  The
prediction  for $d_n$  always   stays below the   current experimental
limit, and the predicted values for $d_\mu$  do not reach the proposed
experimental   sensitivity  for   almost   all  relevant   values   of
$\tan\beta$.   It is amusing to remark  that no EDM constraints can be
imposed on  the  CPX scenario  in the  range: $4 \stackrel{<}{{}_\sim}
\tan\beta    \stackrel{<}{{}_\sim}   12$,  which  is   interesting for
analyzing Higgs-boson searches at  high-energy colliders.  In fact, if
the  gluino phase   is chosen  to  be ${\rm  arg}\, (m_{\tilde{g}})  =
35^\circ$  (see    Fig.~\ref{fig4}(b)),    the  different   EDM  terms
contributing  to $d_{\rm  Tl}$ approximately  cancel and $d_{\rm  Tl}$
does not  exceed much  the experimental  limit.  Similarly, since  the
top/stop- CEDM effects are small in this CPX scenario, the neutron EDM
is always  smaller than  its  conservative experimental   upper bound:
$1.2\times 10^{-25}$~$e$~cm.  However, for $\tan\beta \approx 40$, the
muon EDM  can  be significant,   and  its value  $d_\mu \sim  4.\times
10^{-24}$~$e$~cm lies well within the proposed explorable range.  This
example nicely illustrates the important  r\^ole of complementarity of
a high-sensitivity measurement  of  a  muon  EDM in constraining   the
CP-violating parameter space of the MSSM.

It is also interesting to examine  the dependence of the different EDM
contributions shown  in  Fig.~\ref{fig4} on  the  $\mu$-parameter, for
large values of $\tan\beta$.  In Fig.~\ref{fig5}, we display numerical
values  of $d_{\rm Tl}$, $d_n$ and  $d_\mu$ as functions  of $\mu$ for
two   scenarios  with   $\tan\beta   = 40$,    $M_{\rm SUSY}=  1$~TeV,
$m_{\tilde{g}} = 1$~TeV,   $m_{\widetilde{W}}  = m_{\widetilde{B}}   =
0.3$~TeV, ${\rm arg} (m_{\tilde{g}})  = {\rm arg} (m_{\widetilde{W}} )
= 90^\circ$, $A_{t,b} = 2$~TeV,  ${\rm arg} (A_{t,b}) = 90^\circ$: (a)
$M_{H^+}   = 150$~GeV;  (b)   $M_{H^+} =  300$~GeV.   In  analogy with
Fig.~\ref{fig4}, the individual  contributions to $d_{\rm  Tl}$ due to
top/stop   and  chargino  two-loop  graphs   and due   to the   CP-odd
electron--nucleon    operator      $C_S$   are    also      shown.  In
Fig.~\ref{fig5}(a), we observe   that the different CP-violating   EDM
operators  may cancel in   $d_{\rm Tl}$ and   $d_n$, even for  smaller
values  of  the $\mu$-parameter,  i.e.\ for $\mu  \approx 700$~GeV. In
this region of  parameter space, the  muon EDM is  predicted to be  as
large as $0.8\times 10^{-23}$~$e$~cm, which  falls within the reach of
the proposed $d_\mu$  measurement.    In Fig.~\ref{fig5}(b), we   give
numerical  estimates of $d_{\rm   Tl}$,  $d_n$ and $d_\mu$  for  a CPX
scenario  with a heavier  charged  Higgs boson,  i.e.\ for  $M_{H^+} =
300$~GeV. Again, we find that the predicted value for $d_{\rm Tl}$ can
be close  to the experimental limit  for a wide range of $\mu$-values,
while   $d_\mu$    always   stays  above the   proposed   experimental
sensitivity.

Let us summarize the focal points of  this section. We have explicitly
demonstrated that the non-observation of  the thallium EDM can provide
strict   constraints  on  the    CP-violating parameters  related   to
third-generation   squarks, charginos  and  gluinos.  The  constraints
derived from  the     neutron  EDM   limit   are less     restrictive.
Nevertheless,  our  numerical   analysis  has   also shown   that  the
constraints from  the thallium EDM  can be  significantly weakened, if
the different CP-violating operators $d_e$ and $C_S$ cancel in $d_{\rm
Tl}$.  For instance, this could be the case for the benchmark scenario
CPX,  for   low   and  intermediate  values   of   $\tan\beta$.   Such
cancellations of the CP-violating operators $d_e$  and $C_S$ can occur
for a wide range of  parameters and crucially  depend on the choice of
the phase  combinations: ${\rm arg}\,  (\mu A_t  )$, ${\rm arg}\, (\mu
m_{\tilde{g}})$ and ${\rm arg}\, (\mu m_{\tilde{W}})$.  In particular,
we find that a possible high-sensitivity measurement of $d_\mu$ to the
proposed  level of    $10^{-24}$~$e$~cm can constrain  such  uncovered
ranges  of CP-violating   parameters  in a   rather complementary way.
Finally,  unless $M_{H^+}$ is of the  TeV  order, EDM constraints from
$d_{\rm Tl}$ on  scenarios  favoured by electroweak  baryogenesis  are
rather stringent.   They generally   imply either suppressed  chargino
phases,   i.e.~${\rm   arg}\,(m_{\widetilde{W}}) \stackrel{<}{{}_\sim}
10^\circ$, or modest cancellations in  1 part to  10 with one-loop EDM
terms induced by the first two generations of sleptons.

\setcounter{equation}{0}
\section{Conclusions}

To avoid the known CP and FCNC crises  in the MSSM, we have considered
a  framework,   in which the   first two   generations of  squarks and
sleptons are heavier than $\sim 10$~TeV, while the third generation is
light, with  masses not larger than the  order of  a TeV.  Within this
framework of the MSSM,  we have performed   a systematic study  of the
dominant two- and higher-loop  contributions to the  thallium, neutron
and muon EDMs, which   are induced by $b$-,  $t$-quarks, $\tilde{b}$-,
$\tilde{t}$-squarks,  charginos  and   gluinos. At  present,  the most
severe limits are obtained from the  non-observation of a thallium EDM
$d_{\rm Tl}$, whereas experimental   upper limits on the   neutron EDM
$d_n$ are less  stringent  and usually constrain large   contributions
from  a $d$-quark CEDM   and the CP-odd  three-gluon  operator.  Also,
theoretical  predictions  for  $d_n$   are   plagued by  a number   of
uncertainties while estimating hadronic matrix elements.

The largest effects  on the thallium  EDM $d_{\rm Tl}$ result from two
operators,   the CP-odd electron--nucleon     operator $C_S$ and   the
electron EDM  $d_e$.   These two  CP-violating  operators are formally
induced at the two- and higher-loop levels and involve the exchange of
CP-mixed    Higgs   bosons.    Thus,    strong   constraints    on the
radiatively-generated CP-violating Higgs   sector of the MSSM  can  be
derived   from $d_{\rm  Tl}$, and   hence  on the  analyses for direct
searches  of CP-violating Higgs bosons  at high-energy colliders, such
as  LEP2, Tevatron and LHC~\cite{CEMPW}.     In this context, we  have
analyzed the compatibility of an  earlier suggested benchmark scenario
of  maximal CP violation for LEP2  Higgs studies (CPX)~\cite{CPX} with
the thallium  and neutron  EDMs.  We  have  observed the existence  of
strong correlations  among the different  EDM terms, which  enable the
suppression  of  $d_{\rm  Tl}$ and   $d_n$    even below  the  present
experimental  limits.  Specifically,     for   $4\stackrel{<}{{}_\sim}
\tan\beta \stackrel{<}{{}_\sim} 12$  in the CPX scenario with $M_{H^+}
= 150$~GeV, the stop,  gluino, and chargino  phases are all allowed to
receive their maximal values,   i.e.~${\rm arg}\, (A_t) = {\rm  arg}\,
(m_{\tilde{g}}) = {\rm arg}\, (m_{\widetilde{W}}) = 90^\circ$, without
being in conflict with   EDM limits (cf.\  Fig.~\ref{fig4}(a)).   Most
interestingly,  for specific choices of the  gluino phase, the allowed
range of $\tan\beta$ values compatible with EDM limits can be enlarged
dramatically.   For  instance,    if ${\rm arg}\,    (m_{\tilde{g}}) =
35^\circ$  in   the     aforementioned  CPX    scenario  (see     also
Fig.~\ref{fig4}(b)),    the         predicted     values    for    $25
\stackrel{<}{{}_\sim}  \tan\beta   \stackrel{<}{{}_\sim}  45$  do  not
contradict  upper  limits on  thallium    and neutron  EDMs.  For  the
remaining range of   $\tan\beta$ values, the  obtained prediction does
not exceed the $2\sigma$ upper bound on  $|d_{\rm Tl}|$ by a factor of
$\sim$~3.  Evidently, the degree of cancellations required between the
one- and two-loop EDM terms in the  CPX scenario is not excessive, for
certain choices of the gluino phase.

At this  point, it  is  important to  stress that  a muon EDM  $d_\mu$
measured at the $10^{-24}$~$e$~cm-level will help to sensitively probe
CP-violating  regions of the  MSSM  parameter  space  which cannot  be
accessed easily by measurements of the thallium and neutron EDMs. This
complementarity property  is  mainly a consequence  of   the fact that
$d_\mu$ is free from interfering CP-odd electron--nucleus interactions
thanks to  the $C_S$ operator, which   can contribute significantly to
$d_{\rm  Tl}$.  Unlike the neutron  EDM $d_n$, $d_\mu$ does not suffer
from hadronic uncertainties.  Given   the absence of  a signal  in the
measurements of $|d_{\rm Tl}|$ and $|d_n|$, one may now wonder whether
a positive signal in $d_\mu$ would  already imply a positive signal on
$g-2$ as well.  This is not the case within our framework of the MSSM.
If the first two generations of sfermions  are above the TeV scale, the
biggest  contribution  to  $g-2$  comes  again  from  related two-loop
Barr--Zee-type graphs.  However, for phenomenologically viable charged
Higgs-boson  masses $M_{H^+}  \stackrel{>}{{}_\sim}  120$~GeV  in  the
MSSM~\cite{OPAL}, these  effects  on $g-2$ are negligible~\cite{CCCK}.
Then, only  post-LEP2    high-energy colliders and  the   proposed BNL
experiment~\cite{Yannis}   on the muon  EDM  $d_\mu$  might be able to
sensitively explore the  CP-violating parameter   space of the   above
framework of the MSSM in a rather complementary manner.

We have also studied the impact of EDM constraints on the mechanism of
electroweak  baryogenesis induced by  CP-violating chargino  currents. 
For  this purpose,  we considered  a scenario  in~(\ref{EWbau}), which
favours the above  mechanism of electroweak baryogenesis~\cite{CMQSW}. 
In such a scenario, the chargino two-loop graphs of Fig.~\ref{fedm}(d)
represent the dominant contribution to $d_e$ and $d_{\rm Tl}$ as well.
However, as we detailed  in Section~4, our theoretical predictions for
$d_e$   are   at  variance   with   those   presented   in  a   recent
communication~\cite{CCK}.  Moreover,  we find that  LEP2 direct limits
on  Higgs-boson  masses  require  intermediate and  larger  values  of
$\tan\beta$,   i.e.\  $\tan\beta   \stackrel{>}{{}_\sim}  6$,   for  a
phenomenologically  viable scenario  of electroweak  baryogenesis.  In
this $\tan\beta$  regime, experimental upper limits  on $|d_{\rm Tl}|$
give  rise to  strict  constraints, especially  when no  cancellations
between the chargino  two-loop and one-loop EDM terms  are assumed. In
the  latter case,  the charged  Higgs-boson mass  $M_{H^+}$  should be
relatively  large, i.e.\  $M_{H^+} \stackrel{>}{{}_\sim}  700$~GeV for
$\tan\beta       \stackrel{>}{{}_\sim}        6$       and       ${\rm
  arg}\,(m_{\widetilde{W}})     \stackrel{<}{{}_\sim}    90^\circ$.    
Otherwise, for lighter charged Higgs bosons, either the chargino phase
should be suppressed by a factor  of at least 10 or cancellations in 1
part to 10 with one-loop EDM terms need be invoked.

In our computation  of the   Higgs-boson  loop-induced EDMs,  we  have
considered   resummation   effects of   higher-order CP-conserving and
CP-violating   terms in  Higgs-boson self-energies   and vertices.  In
particular, the original  $t$-quark  two-loop graph suggested  by Barr
and Zee~\cite{BZ}  occurs beyond  the two-loop   approximation through
threshold effects in the  $H_i\bar{t}t$ coupling and, depending on the
choice of  the   gluino  phase,  it  might   even  compete  with   the
$\tilde{t}$-squark two-loop graph~\cite{CKP}.  Since our   resummation
of higher-order terms relied on an  effective Lagrangian approach, one
may worry about the relevance of other higher-order terms present in a
complete  computation.  At this stage,  we can only offer estimates of
those   possible   higher-order  electroweak   uncertainties  in   the
calculation  of EDMs.  Thus, we    have checked our results with   and
without resumming the    Higgs-boson self-energies.  In  this  way, no
large modifications are found in our predictions; the variation of our
results is generally less than 10\% for $M_{H^+} \stackrel{<}{{}_\sim}
170$~GeV, and   becomes even smaller,  to less   than 1\% for $M_{H^+}
\stackrel{>}{{}_\sim} 200$~GeV.  This  may  be attributed to the  fact
that the dominant  contributions to EDMs  come from the heaviest Higgs
bosons,  on  which the  relative impact  of  radiative effects is less
important.  On    the  other  hand,   CP-violating  threshold  effects
constitute the main   source   of theoretical uncertainties   in   the
calculation of the original  Barr--Zee graph of Fig.~\ref{fedm}(c), as
they   are      less    controllable     for      low    values     of
$\tan\beta$.\footnote[5]{A   crude      estimate  suggests  that these
additional higher-order   effects  are smaller  than  20\%.}   In this
context, we remark that even the computation of the CP-odd three-gluon
operator is haunted by relevant higher-order electroweak uncertainties
in  the MSSM~\cite{DDLPD}.  The Weinberg operator  can be generated in
its original fashion~\cite{SW} at three and higher loops which involve
CP-violating self-energy and  vertex  subgraphs of Higgs  bosons.   It
then appears necessary  to   develop improved techniques  that   would
enable us  to provide  accurate  estimates of  (resummed) higher-order
terms in the calculation of EDMs.  The  present work is a step towards
this goal.

\subsection*{Acknowledgements} 
The author thanks Marcela Carena  and Carlos Wagner for discussions on
issues related to electroweak  baryogenesis, Adam Ritz for comments on
the  computation of  the mercury  EDM, Maxim  Pospelov  for clarifying
remarks  pertinent to~\cite{LP},  Darwin Chang  and Wai-Yee  Keung for
communications  with regard  to~\cite{CCK}, and  Athanasios  Dedes for
critical remarks.

\newpage

\def\theequation{\Alph{section}.\arabic{equation}}
\begin{appendix}
\setcounter{equation}{0}
\section{Effective Higgs-boson couplings}

The  couplings   of the CP-mixed  Higgs  bosons   $H_{1,2,3}$ to $t$-,
$b$-quarks, $\tilde{t}$-,  $\tilde{b}$-squarks  and charginos $\chi^+$
play a  key r\^ole in our calculations.   In  this appendix we present
the effective Lagrangians   describing the  above interactions,  after
including dominant one- and two-loop CP-even/CP-odd quantum effects on
the Higgs-boson masses and their respective mixings.

Following  the conventions  of~\cite{CEPW},  we first  write down  the
effective  Lagrangian  of  the   Higgs-boson  couplings  to  top  and
bottom quarks
\begin{equation}
  \label{Hqq}
{\cal L}_{H\bar{q}q}\ =\ - \sum_{i=1}^3\, H_i\,
\bigg[\,\frac{g_w m_b}{2 M_W}\, \bar{b}\,\Big( g^S_{H_ibb}\, +\,
ig^P_{H_ibb}\gamma_5 \Big)\, b\: +\: \frac{g_w m_t}{2 M_W}
\, \bar{t}\,\Big( g^S_{H_itt}\, +\,
ig^P_{H_itt}\gamma_5 \Big)\, t\, \bigg]\, ,
\end{equation}
with~\cite{CEPW}\footnote[6]{Here,  we have  also used  the  fact that
  $b$- and $t$-quark masses are  positive, i.e.\ ${\rm Im}\, m_b \propto
  {\rm Im}\, [ h_b + (\delta h_b)  + (\Delta h_b) \tan\beta ] = 0$ and
  ${\rm Im}\, m_t \propto {\rm Im}\, [ h_t + (\delta h_t) + (\Delta h_t)
  \cot\beta ] = 0$.}
\begin{eqnarray}
  \label{gSHbb}
g^S_{H_ibb} & =& {\rm Re}\, \bigg[\,
\frac{1\, +\, (\delta h_b/h_b) }{1\, +\, (\delta h_b/h_b)\, 
+\, (\Delta h_b/h_b) \tan\beta}\,\bigg]\, \frac{O_{1i}}{\cos\beta}
\nonumber\\
&& +\  {\rm Re}\, \bigg[\, \frac{(\Delta h_b/h_b)}{1\, +\, 
(\delta h_b/h_b)\, +\, (\Delta h_b/h_b) \tan\beta}\,\bigg]\ 
\frac{O_{2i}}{\cos\beta}\nonumber\\
&& +\ {\rm Im}\, \bigg[\,
\frac{(\Delta h_b/h_b)\, (\tan^2\beta\, +\, 1)}{1\, +\, 
(\delta h_b/h_b)\, +\, (\Delta h_b/h_b) \tan\beta}\,\bigg]\
O_{3i}\, , \\[0.5cm]
  \label{gPHbb}
g^P_{H_ibb} & =& -\, {\rm Re}\, \bigg\{\, 
\frac{ [ 1\, +\, (\delta h_b/h_b)]\,\tan\beta\, -\, 
(\Delta h_b/h_b)}{1\, +\, (\delta h_b/h_b)\, +\, 
(\Delta h_b/h_b) \tan\beta}\,\bigg\}\, O_{3i}
\nonumber\\
&& +\  {\rm Im}\, \bigg[\, \frac{(\Delta h_b/h_b)\,\tan\beta}{1\, +\, 
(\delta h_b/h_b)\, +\, (\Delta h_b/h_b) \tan\beta}\,\bigg]\ 
\frac{O_{1i}}{\cos\beta}\nonumber\\
&& -\ {\rm Im}\, \bigg[\,
\frac{(\Delta h_b/h_b)}{1\, +\, 
(\delta h_b/h_b)\, +\, (\Delta h_b/h_b) \tan\beta}\,\bigg]\
\frac{O_{2i}}{\cos\beta}\ , \\[0.5cm]
  \label{gSHtt}
g^S_{H_itt} & =& {\rm Re}\, \bigg[\,
\frac{1\, +\, (\delta h_t/h_t)}{1\, +\, (\delta h_t/h_t)\, 
+\, (\Delta h_t/h_t) \cot\beta}\,\bigg]\, \frac{O_{2i}}{\sin\beta}
\nonumber\\
&& +\  {\rm Re}\, \bigg[\, \frac{(\Delta h_t/h_t)}{1\, +\, 
(\delta h_t/h_t)\, +\, (\Delta h_t/h_t) \cot\beta}\,\bigg]\ 
\frac{O_{1i}}{\sin\beta}\nonumber\\
&& +\ {\rm Im}\, \bigg[\,
\frac{(\Delta h_t/h_t)\, (\cot^2\beta\, +\, 1)}{1\, +\, 
(\delta h_t/h_t)\, +\, (\Delta h_t/h_t) \cot\beta}\,\bigg]\
O_{3i}\, , \\[0.5cm]
  \label{gPHtt}
g^P_{H_itt} & =& -\, {\rm Re}\, \bigg\{\,
\frac{[ 1\, +\, (\delta h_t/h_t) ]\,\cot\beta\, -\, 
(\Delta h_t/h_t)  }{1\, +\, (\delta h_t/h_t)\, 
+\, (\Delta h_t/h_t) \cot\beta}\,\bigg\}\, O_{3i}
\nonumber\\
&& +\  {\rm Im}\, \bigg[\, \frac{(\Delta h_t/h_t)\,\cot\beta}{1\, +\, 
(\delta h_t/h_t)\, +\, (\Delta h_t/h_t) \cot\beta}\,\bigg]\ 
\frac{O_{2i}}{\sin\beta}\nonumber\\
&& -\ {\rm Im}\, \bigg[\,
\frac{(\Delta h_t/h_t)}{1\, +\, 
(\delta h_t/h_t)\, +\, (\Delta h_t/h_t) \cot\beta}\,\bigg]\
\frac{O_{1i}}{\sin\beta}\ .
\end{eqnarray}
In  (\ref{gSHbb})--(\ref{gPHtt}),   $O$ is  a  $3\times 3$-dimensional
mixing matrix that relates weak to mass eigenstates of Higgs bosons in
the   presence      of  CP    violation~\cite{PW,CEPW},    and $\delta
h_{t,b}/h_{t,b}$     and   $\Delta     h_{t,b}/h_{t,b}$      represent
non-logarithmic threshold   contributions to  bottom   and  top Yukawa
couplings~\cite{EMa}.  As is  shown in Fig.~A1, the latter  quantities
are predominantly induced  by   gluino  and Higgsino loops.   In   the
presence of CP violation, their analytic forms are~\cite{CEPW}
\begin{eqnarray}
  \label{dhb}
\frac{\delta h_b}{h_b} &=& 
-\frac{2 \alpha_s}{3\pi} m^*_{\tilde{g}} A_b I(m_{\tilde{b}_1}^2,
m_{\tilde{b}_2}^2,|m_{\tilde{g}}|^2)\ 
-\ \frac{|h_t|^2}{16\pi^2} |\mu|^2
I(m_{\tilde{t}_1}^2,m_{\tilde{t}_2}^2,|\mu|^2)\, ,\\
  \label{Dhb}
\frac{\Delta h_b}{h_b} &=& 
\frac{2 \alpha_s}{3\pi} m^*_{\tilde{g}} \mu^* I(m_{\tilde{b}_1}^2,
m_{\tilde{b}_2}^2,|m_{\tilde{g}}|^2)\ +\ \frac{|h_t|^2}{16\pi^2} A^*_t \mu^*
I(m_{\tilde{t}_1}^2,m_{\tilde{t}_2}^2,|\mu|^2)\, ,\\
  \label{Dht}
\frac{\Delta h_t}{h_t} &=& 
\frac{2 \alpha_s}{3\pi} m^*_{\tilde{g}} \mu^* I(m_{\tilde{t}_1}^2,
m_{\tilde{t}_2}^2,|m_{\tilde{g}}|^2)\ +\ \frac{|h_b|^2}{16\pi^2} 
A_b^* \mu^* I(m_{\tilde{b}_1}^2,m_{\tilde{b}_2}^2,|\mu|^2)\, ,\\
  \label{dht}
\frac{\delta h_t}{h_t} &=& 
-\frac{2 \alpha_s}{3\pi} m^*_{\tilde{g}} A_t I(m_{\tilde{t}_1}^2,
m_{\tilde{t}_2}^2,|m_{\tilde{g}}|^2)\ 
-\ \frac{|h_b|^2}{16\pi^2} |\mu|^2
I(m_{\tilde{b}_1}^2,m_{\tilde{b}_2}^2,|\mu|^2)\, ,
\end{eqnarray}
where  $\alpha_s =   g^2_s/(4\pi)$  is the  SU(3)$_c$ fine   structure
constant, and $I(a,b,c)$ is the one-loop function
\begin{equation}
I(a,b,c)\ =\ \frac{ a b \ln (a/b) + b c \ln (b/c) + a c \ln (c/a)}
{(a-b)(b-c)(a-c)}\ .
\end{equation}
In addition,  the stop and sbottom masses are given by (with $q=t,b$)
\begin{eqnarray}
  \label{mq12}
m^2_{\tilde{q}_1\, (\tilde{q}_2)} \!\!&=&\!\! \frac{1}{2}\, \bigg\{\,
\widetilde{M}^2_Q\: +\: \widetilde{M}^2_q\: +\: 2\,m^2_q\: +\: 
T^q_z\,\cos 2\beta M^2_Z \\
&&\!\! +(-)\, \sqrt{\Big[\, \widetilde{M}^2_Q - \widetilde{M}^2_q +
\cos 2\beta M^2_Z\, (T^q_z - 2 Q_q \sin^2\theta_w)\, \Big]^2\: +\:
4\,m^2_q\,|A_q - R_q \mu^* |^2\, }\ \bigg\}\, ,\nonumber
\end{eqnarray}
where $Q_t\,  (Q_b) = 2/3\, (-1/3)$,  $T^t_z = - T^b_z  = 1/2$, $R_t\,
(R_b)  =   \cot\beta  \  (\tan\beta)$,  and  $\sin^2\theta_w   =  1  -
M^2_W/M^2_Z$.   

%******************************************************************
%%% Feynman graphs for the radiative H1,2bb and H1,2tt couplings
%******************************************************************
\begin{figure}

\begin{center}
\begin{picture}(320,150)(0,0)
\SetWidth{0.8}
 
\ArrowLine(0,70)(30,70)\ArrowLine(30,70)(60,70)
\ArrowLine(60,70)(90,70)\ArrowLine(90,70)(120,70)
\DashArrowArc(60,70)(30,90,180){3}\DashArrowArc(60,70)(30,0,90){3}
\DashArrowLine(60,130)(60,100){3}
\Text(60,70)[]{\boldmath $\times$}
\Text(0,65)[lt]{$b_L,t_L$}\Text(120,65)[rt]{$b_R,t_R$}
\Text(60,65)[t]{$\tilde{g}$}\Text(65,125)[l]{$\Phi^0_{1,2}$}
\Text(37,100)[r]{$\tilde{b}^*_L,\tilde{t}^*_L$}
\Text(83,100)[l]{$\tilde{b}^*_R,\tilde{t}^*_R$}

\Text(60,25)[]{\bf (a)}

\ArrowLine(200,70)(230,70)\ArrowLine(230,70)(260,70)
\ArrowLine(260,70)(290,70)\ArrowLine(290,70)(320,70)
\DashArrowArc(260,70)(30,90,180){3}\DashArrowArc(260,70)(30,0,90){3}
\DashArrowLine(260,130)(260,100){3}
\Text(260,70)[]{\boldmath $\times$}
\Text(200,65)[lt]{$b_L,t_L$}\Text(320,65)[rt]{$b_R,t_R$}
\Text(245,65)[t]{$\tilde{h}^\mp_{2,1}$}
\Text(275,65)[t]{$\tilde{h}^\mp_{1,2}$}
\Text(265,125)[l]{$\Phi^0_{1,2}$}
\Text(237,100)[r]{$\tilde{t}^*_R,\tilde{b}^*_R$}
\Text(283,100)[l]{$\tilde{t}^*_L,\tilde{b}^*_L$}

\Text(260,25)[]{\bf (b)}

\end{picture}
\end{center}
\vspace{-1.cm}
\noindent
Figure A1: {\em Effective one-loop $\Phi^0_{1,2}\bar{b}b$ and
  $\Phi^0_{1,2}\bar{t}t$
couplings, $\delta
h_{b,t}$ and $\Delta h_{b,t}$, generated by the exchange of (a)~gluinos
$\tilde{g}$ and (b)~Higgsinos $\tilde{h}^\pm_{1,2}$.}
\end{figure}

It is  important to  remark here   that only the  CP-violating  vertex
effects on $g^S_{H_ibb}$ and $g^P_{H_i bb}$, which are proportional to
${\rm Im}\,  [  (\Delta h_b/h_b) \tan^2\beta   ]$ in (\ref{gSHbb}) and
(\ref{gPHbb}),    are  enhanced  for   moderately    large values   of
$\tan\beta$,       i.e.\      $20     \stackrel{<}{{}_\sim}  \tan\beta
\stackrel{<}{{}_\sim}  40$.    However,  for   very  large  values  of
$\tan\beta$, i.e.\  $\tan\beta \stackrel{>}{{}_\sim}  40$, there  is a
$1/\tan^2\beta$-dependent  damping  factor   due     to   CP-violating
resummation  effects which  cancels the  $\tan^2\beta$-enhanced factor
mentioned above.  As  a  consequence, in the  large-$\tan\beta$ limit,
the   coupling factors $g^S_{H_ibb}$    and $g^P_{H_i bb}$ approach  a
$\tan\beta$-independent constant.  A related discussion is  also given
in Section~2.

Another important  ingredient for our computation of  two-loop EDMs is
the diagonal  effective couplings of  the Higgs bosons to  scalar top
and bottom quarks. Taking  the CP-violating Higgs-mixing effects into
account,  the   effective  Lagrangian  of   interest  to  us   may  be
conveniently written in the form
\begin{equation}
  \label{Hsqsq}
{\cal L}^{\rm diag}_{H\tilde{q}^*\tilde{q}}\ =\ \sum_{i=1}^3\, H_i\,
\sum\limits_{q=t,b}\, 
\Big[\, v\,\xi^{(H_i)}_q\, \Big( \,
\tilde{q}^*_1\tilde{q}_1\ -\ \tilde{q}^*_2\tilde{q}_2\, \Big)\ +\
v\,\zeta^{(H_i)}_q\, \Big( \, \tilde{q}^*_1\tilde{q}_1\ 
+\ \tilde{q}^*_2\tilde{q}_2\, \Big)\,\Big]\,,
\end{equation}
where 
\begin{eqnarray}
  \label{xiHt}
\xi^{(H_i)}_t \!\!&=&\!\! \frac{2m^2_t}{v^2\,(m^2_{\tilde{t}_2}\: 
-\: m^2_{\tilde{t}_1})}\, \bigg[\, 
{\rm Im}\, (\mu A_t)\, \frac{O_{3i}}{\sin^2\beta}\: -\: 
{\rm Re}\, (\mu X_t)\, \frac{O_{1i}}{\sin\beta}\: +\:
{\rm Re}\,(A^*_t X_t)\, \frac{O_{2i}}{\sin\beta}\, \bigg]\, ,\qquad \\ 
  \label{xiHbb}
\xi^{(H_i)}_b \!\!&=&\!\! \frac{2m^2_b}{v^2\,  
(m^2_{\tilde{b}_2} - m^2_{\tilde{b}_1})}\, \bigg[\,
{\rm Im}\,(\mu A_b)\, \frac{O_{3i}}{\cos^2\beta}\: -\:
{\rm Re}\,(\mu X_b)\, \frac{O_{2i}}{\cos\beta}\: +\: 
{\rm Re}\,(A^*_b X_b)\, \frac{O_{1i}}{\cos\beta}\,\bigg]\, ,\\
  \label{zetaq}
\zeta^{(H_i)}_t \!\!&=&\!\! -\, \frac{2m^2_t}{v^2}\,\frac{
O_{2i}}{\sin\beta}\ +\ {\cal O}(g_w^2,g'^2)\,,\qquad
\zeta^{(H_i)}_b \ =\ -\, \frac{2m^2_b}{v^2}\,\frac{O_{1i}}{\cos\beta}\ 
+\ {\cal O}(g_w^2,g'^2)\,,\qquad
\end{eqnarray}
with  $X_q  =  A_q  -  R_q  \mu^*$  ($q=t,b$).   Although  we  assumed
$m^2_{\tilde{q}_1}  >  m^2_{\tilde{q}_2}$,  the  effective  Lagrangian
(\ref{Hsqsq}) exhibits  the nice feature that it  is fully independent
of the hierarchy of squark masses.

Finally,  we present  the effective  couplings of  the  CP-mixed Higgs
bosons  $H_{1,2,3}$  to  charginos  $\chi^+_{1,2}$~\cite{INch,CDLch}.  
These may be conveniently described by the effective Lagrangian
\begin{equation}
  \label{Lchi}
{\cal L}_{H\chi^+\chi^-}\ =\ -\, \frac{g_w}{2\sqrt{2}}\,
\sum_{i=1}^3\, H_i\, \sum\limits_{j,k=1,2}\, 
\bar{\chi}^+_j\, \Big( \, a_{H_i\chi^-_j\chi^+_k}\ +\ 
b_{H_i\chi^-_j\chi^+_k}\, i\gamma_5\, \Big)\, \chi^+_k\, ,
\end{equation}
where
\begin{eqnarray}
  \label{aHchi}
a_{H_i\chi^-_j\chi^+_k} \!\!&=&\!\! O_{1i}\, \Big(\, 
C^{R*}_{2j}\, C^L_{1k}\: +\: C^R_{2k}\,C^{L*}_{1j}\,\Big)\ +\
O_{2i}\, \Big(\, 
C^{R*}_{1j}\, C^L_{2k}\: +\: C^R_{1k}\,C^{L*}_{2j}\,\Big)\nonumber\\
&&-\, i\, O_{3i}\, \Big[\, \sin\beta\, \Big(\, 
C^{R*}_{2j}\, C^L_{1k}\: -\: C^R_{2k}\,C^{L*}_{1j}\,\Big)\ +\
\cos\beta\, \Big(\, 
C^{R*}_{1j}\, C^L_{2k}\: -\:
C^R_{1k}\,C^{L*}_{2j}\,\Big)\,\Big]\,,\qquad\ \\
  \label{bHchi}
b_{H_i\chi^-_j\chi^+_k} \!\!&=&\!\! i\,O_{1i}\, \Big(\, 
C^{R*}_{2j}\, C^L_{1k}\: -\: C^R_{2k}\,C^{L*}_{1j}\,\Big)\ +\
i\,O_{2i}\, \Big(\, 
C^{R*}_{1j}\, C^L_{2k}\: -\: C^R_{1k}\,C^{L*}_{2j}\,\Big)\nonumber\\
&&+\, O_{3i}\, \Big[\, \sin\beta\, \Big(\, 
C^{R*}_{2j}\, C^L_{1k}\: +\: C^R_{2k}\,C^{L*}_{1j}\,\Big)\ +\
\cos\beta\, \Big(\, 
C^{R*}_{1j}\, C^L_{2k}\: +\: C^R_{1k}\,C^{L*}_{2j}\,\Big)\,\Big]\,.\qquad\
\end{eqnarray}
In the above, $C^R$ and $C^L$ are  $2\times 2$ unitary matrices, which
diagonalize the chargino mass matrix:
\begin{equation}
  \label{MC}
M_C\ = \ \left(\! \begin{array}{cc}
m_{\widetilde{W}} & g_w \langle \phi^{0*}_2\rangle \\
g_w \langle \phi^0_1 \rangle & \mu \end{array} \!\right)\,,
\end{equation} 
with   $\langle  \phi^0_1   \rangle  =   v_1/\sqrt{2}$   and  $\langle
\phi^{0*}_2   \rangle   =   v_2/\sqrt{2}$,  through   the   bi-unitary
transformation
\begin{equation}
  \label{biMC}
C^{R\dagger} M_C\, C^L\ =\ {\rm diag}\, \Big(\, m_{\chi^+_1},\
m_{\chi^+_2}\,\Big)\, .
\end{equation}
In (\ref{biMC}), the chargino mass-eigenvalues are given by 
\begin{eqnarray}
  \label{mchi12}
m_{\chi^+_1 (\chi^+_2)} &=& \frac{1}{2}\, \bigg[\, |m^2_{\widetilde{W}}|\:
+\: |\mu |^2\: +\: 2M^2_W \nonumber\\
&& -\,(+)\ \sqrt{ (|m^2_{\widetilde{W}}| + |\mu |^2 + 2M^2_W )^2 \: -\:
4\,|m_{\widetilde{W}}\mu \, -\, M^2_W \sin 2\beta |^2\, }\ \bigg]\, ,
\end{eqnarray}
while the  analytic expressions for the mixing  matrices $C^{L,R}$ are
quite  lengthy  in the  presence  of CP  violation,  and  will not  be
presented    here;    they   can    be    computed   using    standard
techniques~\cite{INch}.

For completeness, we give the corresponding effective couplings of the
would-be Goldstone boson $G^0$ to charginos $\chi^+_{1,2}$:
\begin{eqnarray}
  \label{G0chi}
a_{G^0\chi^-_j \chi^+_k} &=& 
i\,\cos\beta\, \Big(\, 
C^{R*}_{2j}\, C^L_{1k}\: -\: C^R_{2k}\,C^{L*}_{1j}\,\Big)\ -\ i\,
\sin\beta\, \Big(\, 
C^{R*}_{1j}\, C^L_{2k}\: -\:
C^R_{1k}\,C^{L*}_{2j}\,\Big)\,,\nonumber\\
b_{G^0\chi^-_j \chi^+_k} &=& -\,
\cos\beta\, \Big(\, 
C^{R*}_{2j}\, C^L_{1k}\: +\: C^R_{2k}\,C^{L*}_{1j}\,\Big)\ +\
\sin\beta\, \Big(\, 
C^{R*}_{1j}\, C^L_{2k}\: +\:
C^R_{1k}\,C^{L*}_{2j}\,\Big)\, .
\end{eqnarray}
A non-trivial consistency check for the correctness of our analytic
results is the vanishing of the diagonal scalar couplings of the $G^0$
boson to charginos, i.e.\ $a_{G^0\chi^-_j\chi^+_j} = 0$.

\end{appendix}

%%%%%%%%%%%%%%%%%%%%%%%%%%%%%%%%%%%%%%%%%%%%%%%%%%%%%%%%%%%%%%%%%%%%%
%                  FIGURES ON NUMERICAL ESTIMATES                   %
%%%%%%%%%%%%%%%%%%%%%%%%%%%%%%%%%%%%%%%%%%%%%%%%%%%%%%%%%%%%%%%%%%%%%

\begin{figure}[t]
   \leavevmode
\vspace{-1.7cm} 
\begin{flushleft}
   \epsfxsize=19.0cm
   \epsfysize=20.0cm
     \epsffile[0 0 680 652]{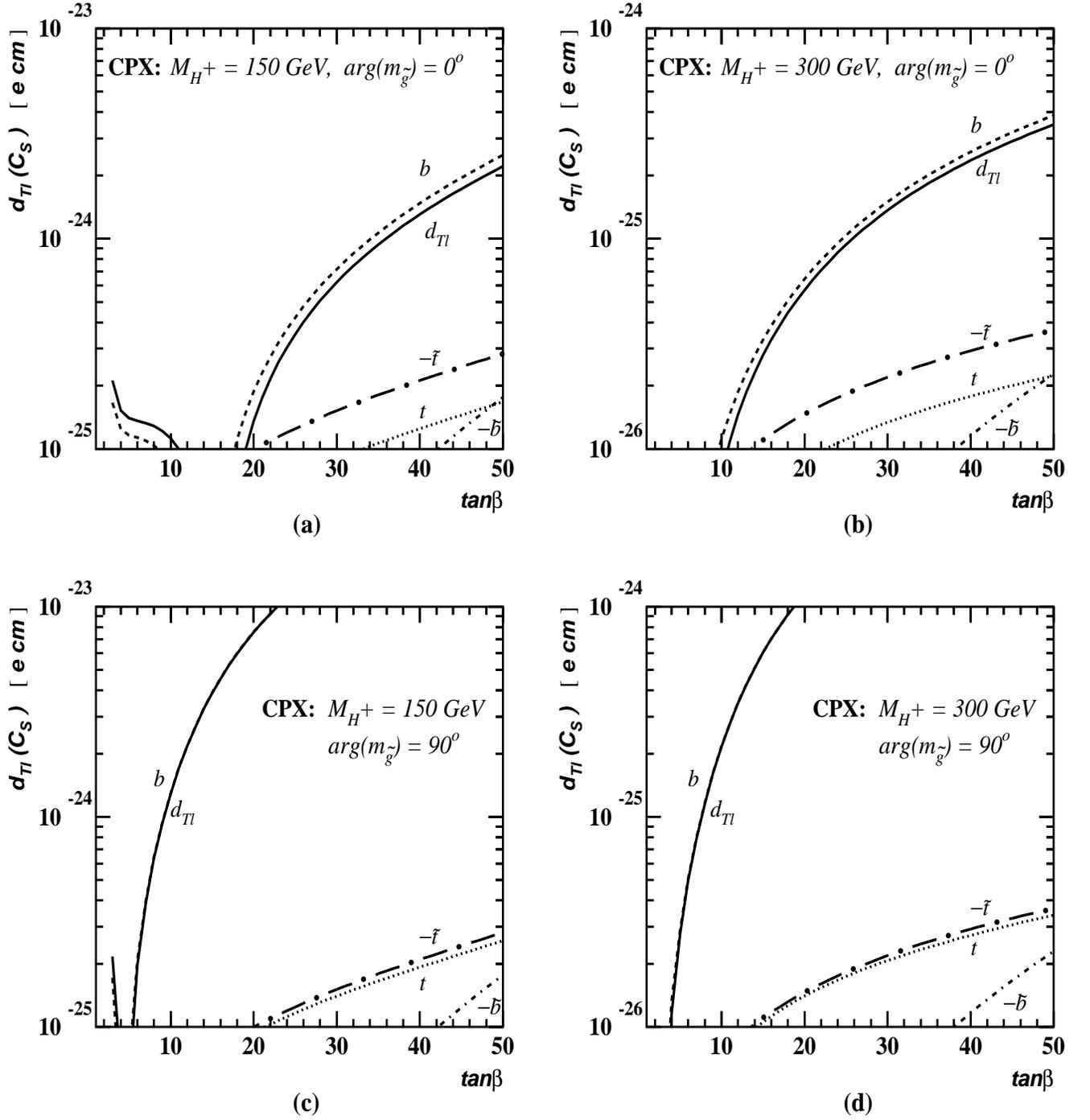}
\end{flushleft}
\vspace{-0.8cm} 
\caption{\em Numerical  estimates   of  $^{205}$Tl EDM  $d_{\rm   Tl}$
  induced by the CP-odd electron--nucleon  operator $C_S$ as functions of
  $\tan\beta$,    in four selected CPX   scenarios  with $M_{\rm SUSY} =
  1$~TeV. The values   of the CPX  parameters are  given in (\ref{CPX}).
  The individual  $b,\tilde{t},t,\tilde{b}$    contributions to  $d_{\rm
  Tl}(C_S)$,      along with   their     relative     signs,   are  also
  displayed.}\label{fig1}
\end{figure}

\begin{figure}
   \leavevmode
\vspace{-2.cm} 
 \begin{center}
   \epsfxsize=14.cm
    \epsffile[0 0 482 652]{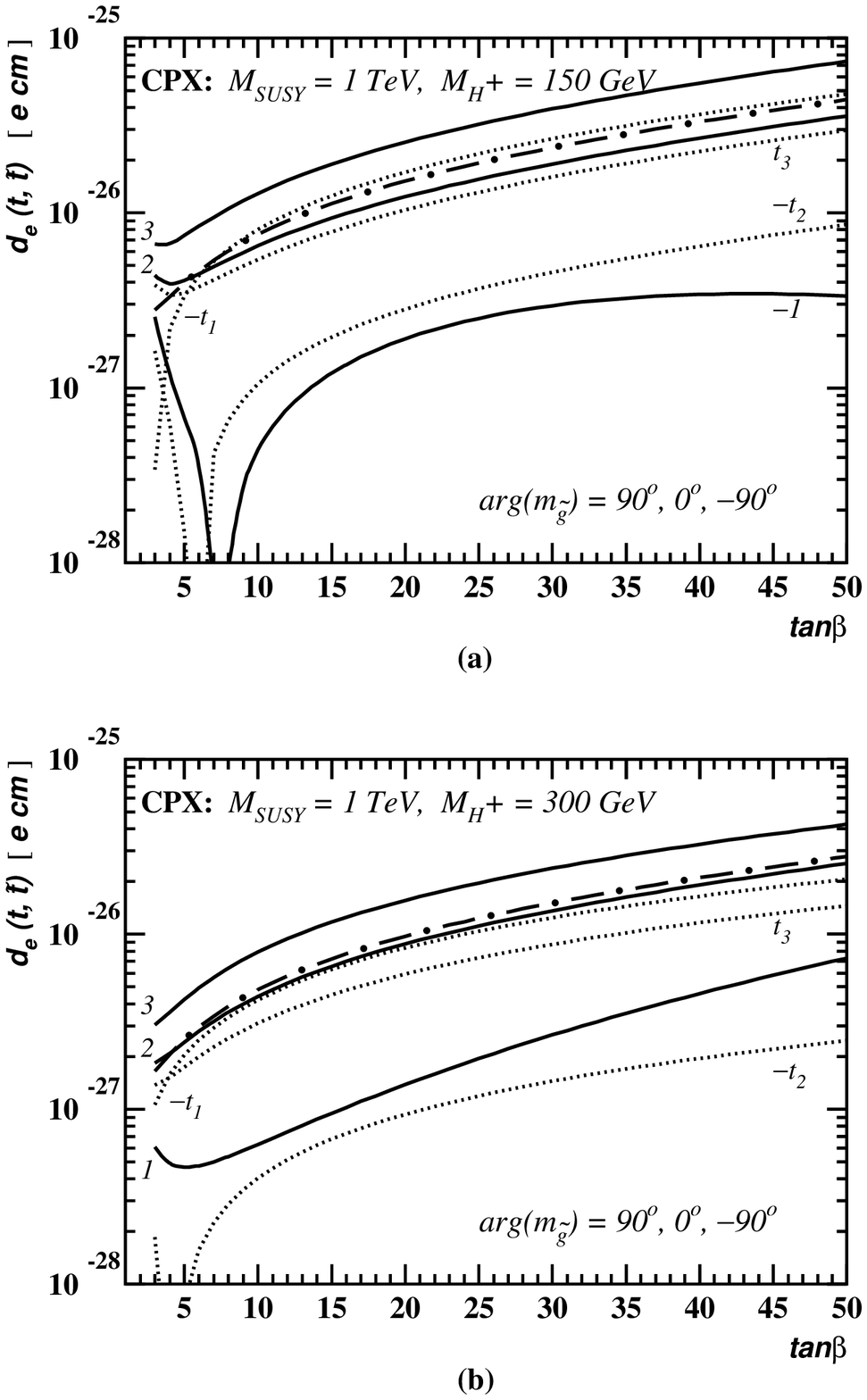}
 \end{center}
\vspace{-0.8cm} 
\caption{\em Numerical estimates of resummed Higgs-boson two-loop 
  effects on $d_e$, induced by $t,b$- quarks and
  $\tilde{t},\tilde{b}$- squarks, as functions of $\tan\beta$, in two
  variants of the CPX scenario, with (a)~$M_{H^+} = 150$~GeV and
  (b)~$M_{H^+} = 300$~GeV.  The long-dash-dotted lines indicate the
  stop/sbottom contributions to $d_e$.  The dotted lines $t_{1,2,3}$
  correspond to top/bottom contributions, for ${\rm arg}
  (m_{\tilde{g}} ) = 90^\circ,\ 0^\circ,\ -90^\circ$, respectively.
  Likewise, the solid lines $1,2,3$ give the sum of all the
  aforementioned contributions to $d_e$ for the same values of gluino
  phases.  Contributions to $d_e$ that are denoted with a minus
  sign are negative.}\label{fig2}
\end{figure}

\begin{figure}
   \leavevmode
\vspace{-2.cm} 
 \begin{center}
   \epsfxsize=14.cm
    \epsffile[0 0 482 652]{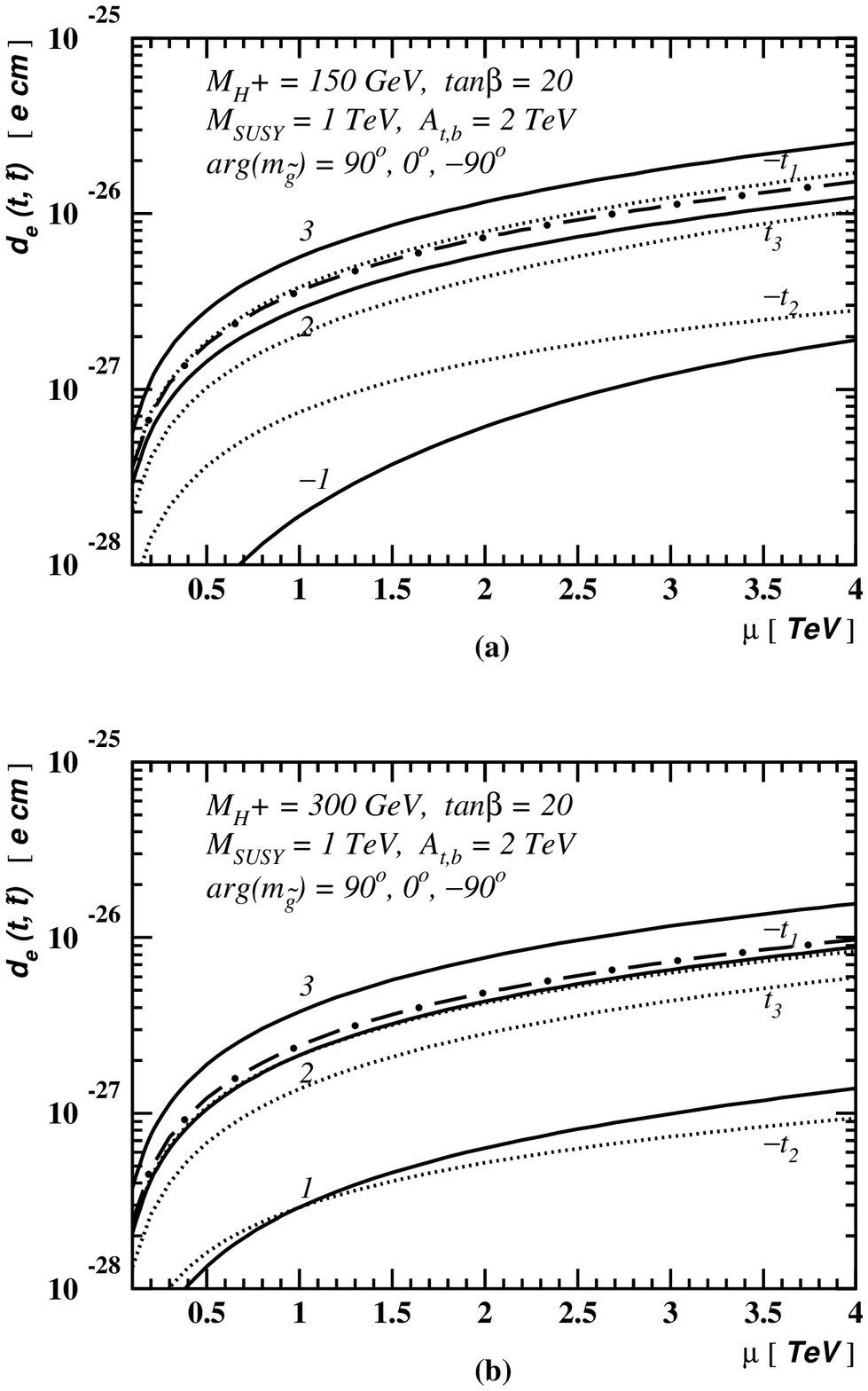}
 \end{center}
\vspace{-1.cm} 
\caption{\em Numerical values of resummed Higgs-boson two-loop effects
  on $d_e$, induced by $t,b$- quarks and $\tilde{t},\tilde{b}$- squarks,
  as functions of $\mu$, in two variants of the CPX scenario, with
  $\tan\beta = 20$, and (a)~$M_{H^+} = 150$~GeV and (b)~$M_{H^+} =
  300$~GeV.  The meaning of the different line types is identical to
  that of Fig.~\ref{fig2}. For $A_{t,b} = 0$, the long-dash-dotted line
  disappears and so the solid lines collapse to the dotted
  ones.}\label{fig2new}
\end{figure}

\begin{figure}
   \leavevmode
\vspace{-2.cm} 
 \begin{center}
   \epsfxsize=14.0cm
    \epsffile[0 0 482 652]{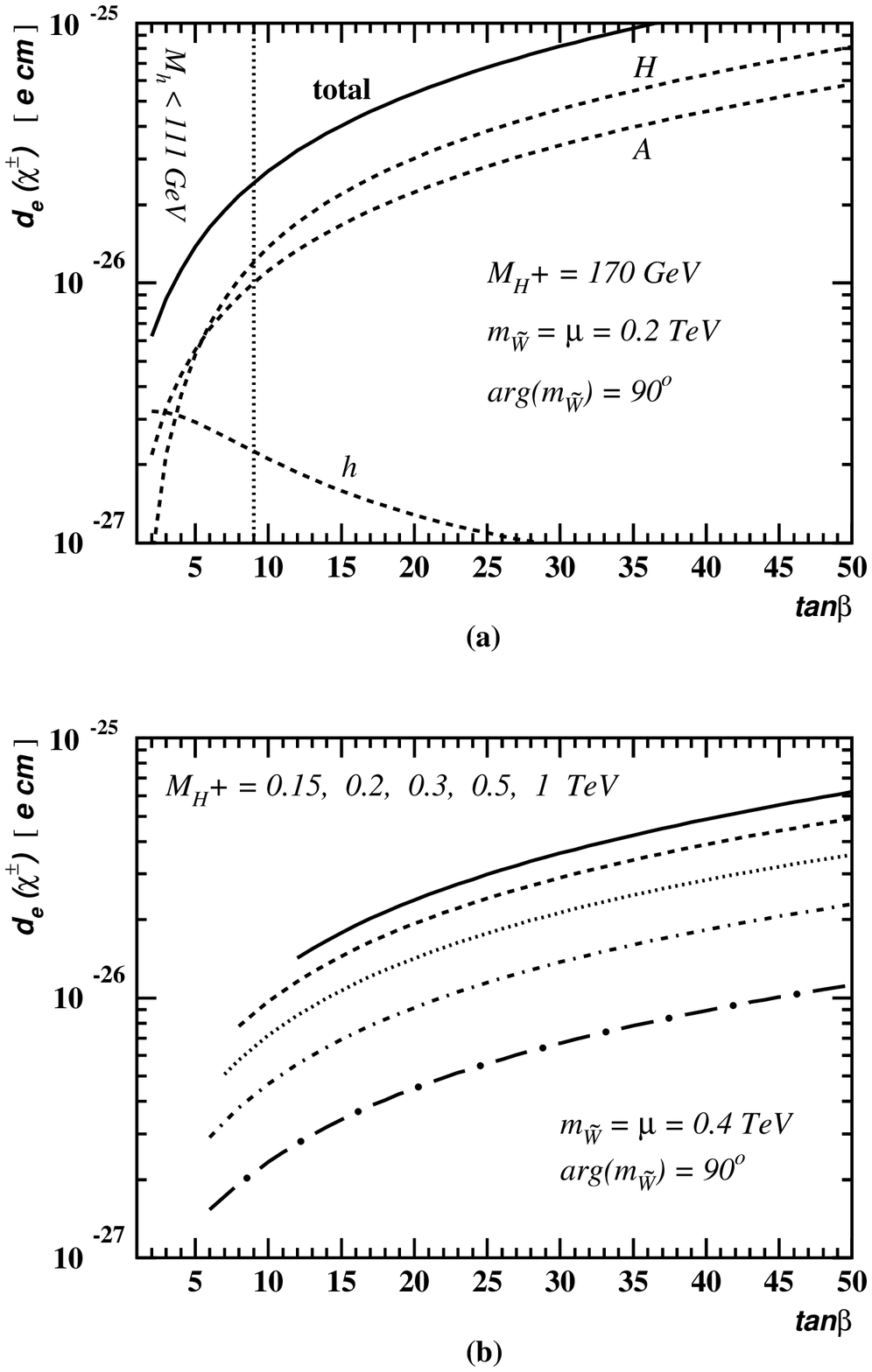}
 \end{center}
 \vspace{-1.cm} 
\caption{\em $d_e$ versus $\tan\beta$ in a scenario favoured by
electroweak baryogenesis, with MSSM parameters $\widetilde{M}_Q =
\widetilde{M}_D = 3$~TeV, $\widetilde{M}_U = 0$, $A_{t,b} = 1.8$~TeV,
$m_{\tilde{g}} = 3$~TeV and ${\rm arg} (A_{t,b} ) = {\rm arg}
(m_{\tilde{g}}) = 0^\circ$.  In (a), $M_{H^+} = 170$~GeV is used,
corresponding to $M_{\mbox{\scriptsize `$A$'}} \approx 150$~GeV, and
$m_{\widetilde{W}} = \mu = 0.2$~TeV and ${\rm arg} (m_{\widetilde{W}})
= 90^\circ$.  Also displayed are the individual `$h$', `$H$', `$A$'
contributions to $d_e$ and the LEP excluded region from direct
Higgs-boson searches.  In (b), numerical values are shown for $M_{H^+}
= 150$~GeV~(solid), 200~GeV~(dashed), 300~GeV~(dotted),
500~GeV~(dash-dotted) and 1~TeV~(long-dash-dotted), in a scenario with
$m_{\widetilde{W}} = \mu = 0.4$~TeV and ${\rm arg} (m_{\widetilde{W}})
= 90^\circ$.}\label{fig3}
\end{figure}

\begin{figure}
   \leavevmode
\vspace{-2.cm}
 \begin{center}
   \epsfxsize=15.0cm
    \epsffile[0 0 482 652]{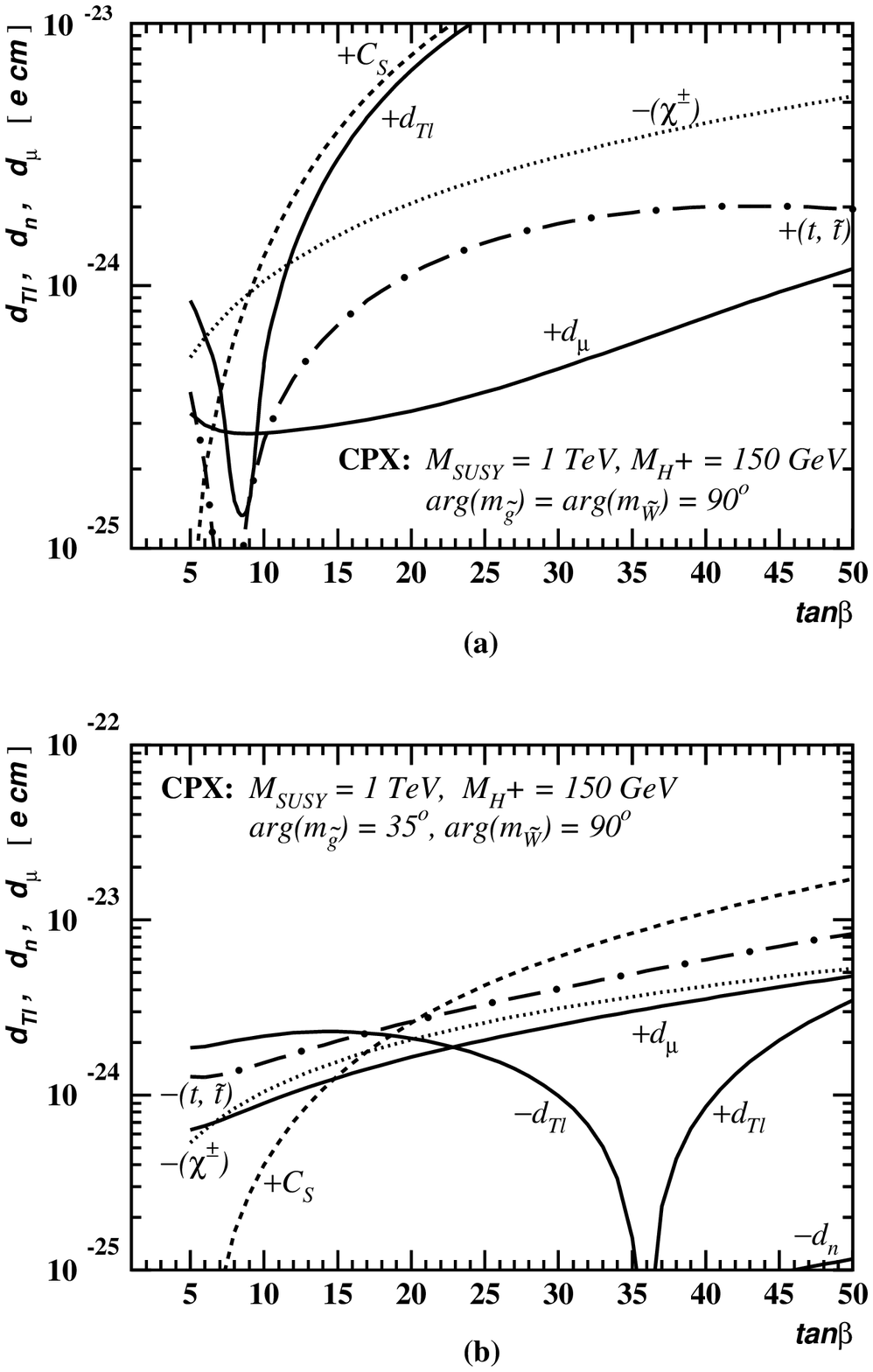}
 \end{center}
 \vspace{-1.cm} 
\caption{\em EDMs of $d_{\rm Tl}$, $d_n$ and $d_\mu$ as functions of
  $\tan\beta$ in two versions of the CPX scenario: (a) ${\rm arg}
  (m_{\tilde{g}}) = {\rm arg} (m_{\widetilde{W}} ) = 90^\circ$, and (b)
  ${\rm arg} (m_{\tilde{g}}) = 35^\circ$, ${\rm arg} (m_{\widetilde{W}} )
  = 90^\circ$. Also shown are the different contributions, along with
  their relative signs, to $d_{\rm Tl}$ from top/stop
  (long-dash-dotted) and chargino (dotted) Higgs-boson two-loop
  graphs, as well as from the CP-odd electron--nucleon coupling $C_S$
  (dashed).}\label{fig4}
\end{figure}

\begin{figure}
   \leavevmode
\vspace{-2.cm}
 \begin{center}
   \epsfxsize=15.0cm
    \epsffile[0 0 482 652]{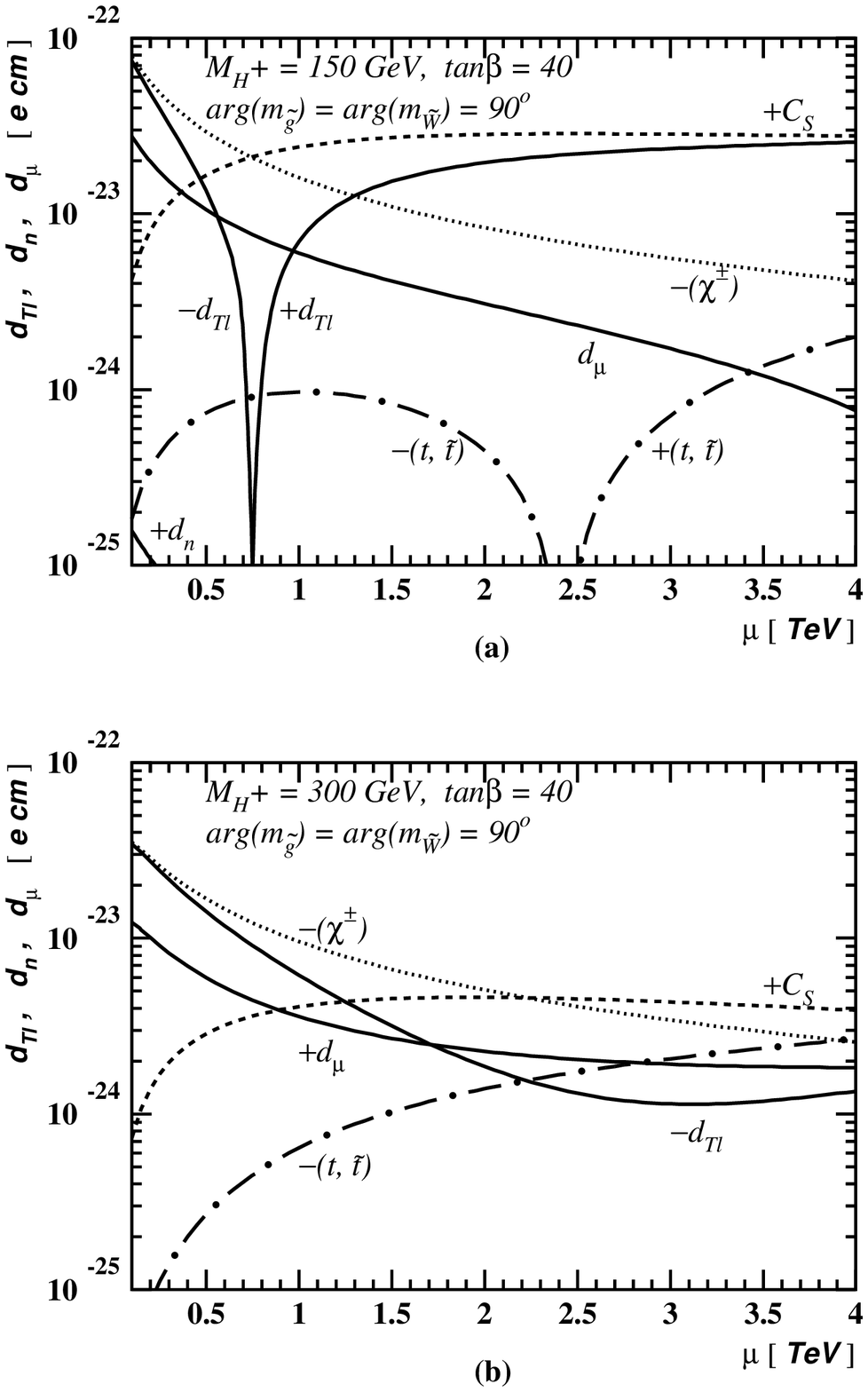}
 \end{center}
 \vspace{-1.cm} 
\caption{\em Numerical values of $d_{\rm Tl}$, $d_n$ and $d_\mu$ as
  functions of $\mu$ for two large-$\tan\beta$ scenarios, with
  $\tan\beta = 40$, $M_{\rm SUSY}= 1$~TeV, $m_{\tilde{g}} = 1$~TeV,
  $m_{\widetilde{W}} = m_{\widetilde{B}} = 0.3$~TeV, ${\rm arg}
  (m_{\tilde{g}}) = {\rm arg} (m_{\widetilde{W}} ) = 90^\circ$, $A_{t,b} =
  2$~TeV, ${\rm arg} (A_{t,b}) = 90^\circ$: (a) $M_{H^+} = 150$~GeV;
  (b) $M_{H^+} = 300$~GeV.  In analogy with Fig.~\ref{fig4}, the
  individual contributions to $d_{\rm Tl}$ due to top/stop and
  chargino two loop graphs and due to the $C_S$ operator are also
  shown.}\label{fig5}
\end{figure}


\newpage


\begin{thebibliography}{99}
  
\bibitem{GD} G.F. Giudice and  S. Dimopoulos, Phys.\ Lett.\ {\bf B357}
  (1995) 573;  G. Dvali and  A. Pomarol,  Phys.\ Rev.\  Lett.\  {\bf 77}
  (1996) 3728; A.G.  Cohen, D.B. Kaplan  and A.E. Nelson, Phys.\ Lett.\
  {\bf B388} (1996) 588; P. Bin\'etruy  and E. Dudas, Phys.\ Lett.\ {\bf
  B389} (1996) 503.

\bibitem{CKP} D. Chang, W.-Y. Keung and A. Pilaftsis, Phys.\ Rev.\ 
  Lett.\ {\bf 82} (1999) 900 and {\bf 83} (1999) 3972 (E).

\bibitem{APedm} A. Pilaftsis, Phys.\ Lett.\  {\bf  B471} (1999) 174;
  D. Chang, W.-F. Chang and W.-Y. Keung, Phys.\ Lett.\ {\bf B478}
  (2000) 239.

\bibitem{BZ} S.M. Barr and A. Zee,  Phys.\ Rev.\  Lett.\  {\bf 65}
  (1990) 21.

\bibitem{APLB} A. Pilaftsis, Phys.\  Rev.\ {\bf D58} (1998) 096010 and
  Phys.\ Lett.\  {\bf B435} (1998)  88.
  
\bibitem{PW} A. Pilaftsis and C.E.M.  Wagner, Nucl.\ Phys.\ {\bf B553}
  (1999) 3; D.A. Demir, Phys.\ Rev.\ {\bf D60} (1999) 055006; S.Y.
  Choi, M. Drees and J.S. Lee, Phys.\ Lett.\ {\bf B481} (2000) 57;
  G.L. Kane and L.-T. Wang, Phys.\ Lett.\ {\bf B488} (2000) 383; S.Y.
  Choi, K. Hagiwara and J.S. Lee, Phys.\ Rev.\ {\bf D64} (2001)
  032004; Phys.\ Lett.\ {\bf B529} (2002) 212, S. Heinemeyer, Eur.\ 
  Phys.\ J. {\bf C22} (2001) 521; T. Ibrahim and P. Nath,
  hep-ph/0204092; S.W. Ham, S.K. Oh, E.J. Yoo, C.M. Kim and D. Son,
  hep-ph/0205244.

\bibitem{INch} T. Ibrahim and P. Nath, Phys.\ Rev.\ {\bf D63} (2001)
  035009.

\bibitem{CEPW} M. Carena, J. Ellis, A. Pilaftsis and C.E.M. Wagner,
  Nucl.\ Phys.\ {\bf B586} (2000) 92.

\bibitem{CPX}  M. Carena, J.  Ellis, A.  Pilaftsis and  C.E.M. Wagner,
  Phys.\ Lett.\ {\bf B495} (2000) 155.

\bibitem{CEPW2} M. Carena, J. Ellis, A. Pilaftsis and C.E.M. Wagner,
  Nucl.\ Phys.\ {\bf B625} (2002) 345.
  
\bibitem{CPHig} For phenomenological applications of radiative
Higgs-sector CP violation in the MSSM, see: S.Y.  Choi and M.  Drees,
Phys.\ Rev.\ Lett.\ {\bf 81} (1998) 5509; S.  Baek and P.  Ko, Phys.\
Rev.\ Lett.\ {\bf 83} (1999) 488; W.  Bernreuther, A.  Brandenburg and
M.  Flesch, hep-ph/9812387; K.S. Babu, C.  Kolda, J.  March-Russell
and F. Wilczek, Phys.\ Rev.\ {\bf D59} (1999) 016004; B.  Grzadkowski,
J.F.  Gunion and J.  Kalinowski, Phys.\ Rev.\ {\bf D60} (1999) 075011;
S.Y.  Choi, M.  Guchait, H.S.  Song and W.Y.  Song, Phys.\ Lett.\ {\bf
B483} (2000) 168; P. Osland, Acta Phys.\ Polon.\ {\bf B30} (1999)
1967; T.  Han, T.  Huang, Z.H.  Lin, J.X.  Wang and X.  Zhang, Phys.\
Rev.\ {\bf D61} (2000) 015006; B.  Grzadkowski and J.  Pliszka, Phys.\
Rev.\ {\bf D60} (1999) 115018; S.Y.  Choi and J.S.  Lee, Phys.\ Rev.\
{\bf D61} (2000) 015003; C.-S.  Huang and L.  Wei, Phys.\ Rev.\ {\bf
D61} (2000) 116002; K.  Freese and P.  Gondolo, hep-ph/9908390;
S.Y. Choi and J.S.  Lee, Phys.\ Rev.\ {\bf D61} (2000) 115002; A.
Dedes and S.  Moretti, Phys.\ Rev.\ Lett.\ {\bf 84} (2000) 22; Nucl.\
Phys.\ {\bf B576} (2000) 29; M.S. Berger, Phys.\ Rev.\ Lett.\ {\bf 87}
(2001) 131801; D.A. Demir, Nucl.\ Phys.\ Proc.\ Suppl.\ {\bf 101}
(2001) 431; S.W. Ham, S.K. Oh, E.J. Yoo and H.K. Lee, J. Phys.\ {\bf
G27} (2001) 1; A.G. Akeroyd and A. Arhrib, Phys.\ Rev.\ {\bf D64}
(2001) 095018; M. Boz and N.K. Pak, Phys.\ Rev.\ {\bf D65} (2002)
075014; A.  Arhrib, D.K. Ghosh and O.C.W. Kong, Phys.\ Lett.\ {\bf
B537} (2002) 217; E.  Christova, S. Fichtinger, S. Kraml and W.
Majerotto, Phys.\ Rev.\ {\bf D65} (2002) 094002; hep-ph/0205227;
C.-H. Chen, hep-ph/0206143; M.N. Dubinin and A.V. Semenov,
hep-ph/0206205; A.  Bartl, K.  Hidaka, T. Kernreiter and W. Porod,
hep-ph/0207186.
 
\bibitem{KRS} V.A. Kuzmin, V.A. Rubakov and M.E. Shaposhnikov, Phys.\
  Lett.\ {\bf B155} (1985) 36.
    
\bibitem{CQW} For recent studies, see: M. Carena, M. Quir\'os and
  C.E.M.  Wagner, Nucl.\ Phys.\ {\bf B524} (1998) 3; M. Laine and K.
  Rummukainen, Phys.\ Rev.\ Lett.\ {\bf 80} (1998) 5259; Nucl.\ Phys.\ 
  {\bf B535} (1998) 423; Nucl.\ Phys.\ {\bf B597} (2001) 23; K.
  Funakubo, Prog.\ Theor.\ Phys.\ {\bf 102} (1999) 389; J. Grant and
  M. Hindmarsh, Phys.\ Rev.\ {\bf D59} (1999) 116014; M. Losada,
  Nucl.\ Phys.\ {\bf B537} (1999) 3; Nucl.\ Phys.\ {\bf B569} (2000)
  125; S.  Davidson, M. Losada and A.  Riotto, Phys.\ Rev.\ Lett.\ 
  {\bf 84} (2000) 4284; A.B. Lahanas, V.C.  Spanos and V. Zarikas,
  Phys.\ Lett.\ {\bf B472} (2000) 119; N. Rius and V. Sanz, Nucl.\ 
  Phys.\ {\bf B570} (2000) 155; J.M.  Cline, M. Joyce and K.
  Kainulainen, JHEP {\bf 0007} (2000) 018; [Erratum:hep-ph/0110031];
  S.J. Huber, P. John and M.G. Schmidt, Eur.\ Phys.\ J. {\bf C20}
  (2001) 695; H. Murayama and A. Pierce, hep-ph/0201261.
  
\bibitem{CMQSW} M. Carena, J.M. Moreno, M. Quir\'os, M. Seco and
  C.E.M. Wagner, Nucl.\ Phys.\ {\bf B599} (2001) 158; and work in
  progress.

\bibitem{EDMexp} P.G. Harris et al., Phys.\ Rev.\ Lett.\ {\bf 82}
  (1999) 904.

\bibitem{EDMTl} B.C. Regan, E.D. Commins, C.J. Schmidt and D. DeMille,
  Phys.\ Rev.\ Lett.\ {\bf 88} (2002) 071805.

\bibitem{SBarr}  S. Barr, Phys.\  Rev.\ Lett.\  {\bf 68}  (1992) 1822;
  Int.\ J. Mod.\ Phys.\ {\bf A8} (1993) 209.

\bibitem{FPT} W. Fischler, S. Paban and S. Thomas, Phys.\ Lett.\
  {\bf B289} (1992) 373.

\bibitem{KL} I.B. Khriplovich and S.K. Lamoreaux, {\em CP Violation
  Without Strangeness} (Springer, New York, 1997).

\bibitem{Yannis} Y.K. Semertzidis et~al., Letter of Intent to BNL
  (Spring of 2000, and Fall of 1997); ``Sensitive Search for a
  Permanent Muon Electric Dipole Moment'' [hep-ph/0012087]. For recent
  theoretical studies, see K.S. Babu, B.  Dutta and R.N. Mohapatra,
  Phys.\ Rev.\ Lett.\ {\bf 85} (2000) 5064; T. Ibrahim and P. Nath,
  Phys.\ Rev.\ {\bf D64} (2001) 093002; J.L. Feng, K.T. Matchev and Y.
  Shadmi, Nucl.\ Phys.\ {\bf B613} (2001) 366; J.R. Ellis, J.  Hisano,
  M.  Raidal and Y. Shimizu, Phys.\ Lett.\ {\bf B528} (2002) 86.

\bibitem{dn90} K.F. Smith et~al., Phys.\ Lett.\ {\bf B234} (1990) 191.

\bibitem{dn99} P.G. Harris et~al., Phys.\ Rev.\ Lett.\ {\bf 82} (1999)
  904.

\bibitem{LG} S.K. Lamoreaux and  R. Golub, Phys.\ Rev.\ {\bf D61}
  (2000) 051301.

\bibitem{PR} M. Pospelov and A. Ritz, Phys.\ Rev.\ Lett.\ {\bf 83}
  (1999) 2526; Nucl.\ Phys.\ {\bf B573} (2000) 177.

\bibitem{AKL} S.A. Abel, S. Khalil and O. Lebedev, Nucl.\ Phys.\ {\bf
    B606} (2001) 151.

\bibitem{dHg} M.V. Romalis, W.C. Griffith and E.N. Fortson, Phys.\
  Rev.\ Lett.\ {\bf 86} (2001) 2505.

\bibitem{FOPR} T. Falk, K.A. Olive, M. Pospelov and R. Roiban, Nucl.\
  Phys.\ {\bf B60} (1999) 3.

\bibitem{Ritz} A. Ritz, private communication.

\bibitem{SW} S. Weinberg, Phys.\ Rev.\ Lett.\ {\bf 63} (1989) 2333.

\bibitem{DDLPD} J. Dai, H.  Dykstra, R.G. Leigh, S. Paban, and D.A.
  Dicus, Phys.\ Lett.\ {\bf B237} (1990) 216 and {\bf B242} (1990) 547
  (E).
  
\bibitem{IN} T.  Ibrahim and P.  Nath, Phys.\ Lett.\ {\bf B418} (1998)
  98, Phys.\ Rev.\ {\bf D57} (1998) 478, {\bf D58} (1998) 019901 (E),
  Phys.\ Rev.\ {\bf D58} (1998) 111301; Phys.\ Rev.\ {\bf D61} (2000)
  093004; M.  Brhlik, G.J.  Good and G.L. Kane, Phys.\ Rev.\ {\bf D59}
  (1999) 115004; M.  Brhlik, L.  Everett, G.L.  Kane and J. Lykken,
  Phys.\ Rev.\ Lett.\ {\bf 83} (1999) 2124; Phys.\ Rev.\ {\bf D62}
  (2000) 035005; S. Abel, S. Khalil and O. Lebedev, Phys.\ Rev.\ 
  Lett.\ {\bf 86} (2001) 5850.
  
\bibitem{EFN} J. Ellis, S.  Ferrara and D.V. Nanopoulos, Phys.\ Lett.\ 
  {\bf B114} (1982) 231; W.  Buchm\"uller and D. Wyler, Phys.\ Lett.\ 
  {\bf B121} (1983) 321; J. Polchinski and M.  Wise, Phys.\ Lett.\ 
  {\bf B125} (1983) 393; F. del Aguila, M.  Gavela, J.  Grifols and A.
  Mendez, Phys.\ Lett.\ {\bf B126} (1983) 71; D.V. Nanopoulos and M.
  Srednicki, Phys.\ Lett.\ {\bf B128} (1983) 61; T. Falk, K.A.  Olive
  and M.  Srednicki, Phys.\ Lett.\ {\bf B354} (1995) 99; M.  Dugan, B.
  Grinstein and L. Hall, Nucl.\ Phys.\ {\bf B255} (1985) 413; R.
  Garisto and J.D.  Wells, Phys.\ Rev.\ {\bf D55} (1997) 1611.
  
\bibitem{PN} P.  Nath, Phys.\ Rev.\   Lett.\ {\bf 66}  (1991) 2565; Y.
  Kizukuri and N. Oshimo, Phys.\ Rev.\ {\bf D46} (1992) 3025.
  
\bibitem{EDMrecent} For recent analyses  of one-loop EDM effects, see
  S.  Pokorski,  J.  Rosiek and  C.A. Savoy, Nucl.\ Phys.\  {\bf B570}
  (2000) 81;  E.  Accomando, R.  Arnowitt and  B. Dutta,  Phys.\ Rev.\ 
  {\bf D61} (2000) 115003; A. Bartl, T. Gajdosik, W. Porod,
  P. Stockinger and H. Stremnitzer, Phys.\ Rev.\ {\bf D60} (1999) 073003.
  
\bibitem{TI} In this respect, our analysis of EDM constraints on a
  CP-violating MSSM Higgs sector is complementary to the one presented
  by T.  Ibrahim, Phys.\ Rev.\ {\bf D64} (2001) 035009.

\bibitem{APRD} A. Pilaftsis, Phys.\ Rev.\ {\bf D62} (2000) 016007. 

\bibitem{THW} T.H. West, Phys.\ Rev.\ {\bf D50} (1994) 7; 
  T. Kadoyoshi and N. Oshimo, Phys.\ Rev.\ {\bf D55} (1997) 1481.
  
\bibitem{CPH} In  our numerical analysis,  we employ the  Fortran code
  {\tt cph+.f} available at {\tt http://home.cern.ch/p/pilaftsi/www/}.
  The code  is based on a recent  RG-improved diagrammatic calculation
  of   Higgs-boson  pole  masses   in  the   MSSM  with   explicit  CP
  violation~\cite{CEPW2}.
  
\bibitem{Mhiggs} For recent two-loop studies of an effective
  CP-conserving Higgs potential in the MSSM, see: M. Carena, H.E.
  Haber, S.  Heinemeyer, W.  Hollik, C.E.M. Wagner and G. Weiglein,
  Nucl.\ Phys.\ {\bf B580} (2000) 29; J.R. Espinosa and R.J. Zhang,
  JHEP {\bf 0003} (2000) 026; M. Carena, S. Mrenna and C.E.M. Wagner,
  Phys.\ Rev.\ {\bf D62} (2000) 055008; J.R. Espinosa and I.  Navarro,
  Nucl.\ Phys.\ {\bf B615} (2001) 82; A. Brignole, G. Degrassi, P.
  Slavich and F. Zwirner, Nucl.\ Phys.\ {\bf B631} (2002) 195;
  hep-ph/0206101; S.P. Martin, hep-ph/0206136.

\bibitem{LP} O. Lebedev and M. Pospelov, hep-ph/0204359.

\bibitem{CCK} D. Chang, W.-F. Chang and W.-Y. Keung, hep-ph/0205084
  [v2, 4~June 2002].

\bibitem{SVZ} M.A. Shifman, A.L. Vainshtein and V.I. Zakharov, Phys.\
  Lett.\ {\bf B78} (1978) 443; see also J. Ellis, M.K. Gaillard and
  D.V. Nanopoulos, Nucl.\ Phys.\ {\bf B106} (1976) 292; S. Dawson and
  H.E. Haber, Int.\ J. Mod.\ Phys.\ {\bf A7} (1992) 107.

\bibitem{OPAL} OPAL Collaboration, ``Interpretation of the Search for
  Neutral Higgs Bosons from $\sqrt{s} = 91$~GeV to $\sqrt{s} =
  209$~GeV in a CP-Violating MSSM Scenario,'' OPAL Physics Note PN505,
  available from {\tt http://opal.web.cern.ch/Opal/pubs/ 
  physnote/html/pn505.html}.

\bibitem{CEMPW} M. Carena, J. Ellis, S. Mrenna, A. Pilaftsis and
  C.E.M. Wagner, work in progress. 
  
\bibitem{CCCK} D.  Chang, W.-F. Chang, C.-H. Chou and W.-Y. Keung,
  Phys.\ Rev.\ {\bf D63} (2001) 091301; A. Dedes and H.E. Haber, JHEP
  {\bf 0105} (2001) 006.
  
\bibitem{EMa} E. Ma, Phys.\ Rev.\ {\bf D39} (1989) 1922; R. Hempfling,
  Phys.\ Rev.\ {\bf D49} (1994) 6168; L.J. Hall, R. Rattazzi and U.
  Sarid, Phys.\ Rev.\ {\bf D50} (1994) 7048; T. Blazek, S. Raby and S.
  Pokorski, Phys.\ Rev.\ {\bf D52} (1995) 4151; M. Carena, M.
  Olechowski, S. Pokorski and C.E.M. Wagner, Nucl.\ Phys.\ {\bf B426}
  (1994) 269; S. Heinemeyer, W. Hollik and G. Weiglein, Phys.\ Lett.\ 
  {\bf B440} (1998) 296; Eur.\ Phys.\ J. {\bf C9} (1999) 343; F.
  Borzumatti, G. Farrar, N. Polonsky and S. Thomas, Nucl.\ Phys.\ {\bf
    B555} (1999) 53; J.R. Espinosa and R.J. Zhang, Nucl.\ Phys.\ {\bf
    B586} (2000) 3.

\bibitem{CDLch} S.Y. Choi, M. Drees, J.S. Lee and J. Song,
  hep-ph/0204200.


\end{thebibliography}
\end{document}